\documentclass[%
 aip,
 rsi,
 amsmath,amssymb,
 reprint,%
]{revtex4-1}

\usepackage{graphicx}
\usepackage{dcolumn}
\usepackage{bm}
\usepackage{comment}
\usepackage{braket}
\usepackage{booktabs}

\usepackage[export]{adjustbox}

\usepackage{graphicx}
\usepackage{dcolumn}
\usepackage{bm}

\usepackage[utf8]{inputenc}
\usepackage[T1]{fontenc}
\usepackage{mathptmx}
\usepackage{etoolbox}

\usepackage{color}

\makeatletter
\def\@email#1#2{%
 \endgroup
 \patchcmd{\titleblock@produce}
  {\frontmatter@RRAPformat}
  {\frontmatter@RRAPformat{\produce@RRAP{*#1\href{mailto:#2}{#2}}}\frontmatter@RRAPformat}
  {}{}
}%
\makeatother
\begin{document}

\preprint{AIP/123-QED}
\title{Nanophotonic Phased Array XY Hamiltonian Solver }

\author{Michelle Chalupnik}
\email{mvchalupnik@gmail.com, mo.soltani@rtx.com}
\affiliation{Raytheon BBN Technologies, Cambridge, MA 02138, USA}
\affiliation{Department of Physics, Harvard University, Cambridge, MA 02138, USA}
\affiliation{Currently with Aliro Quantum, Brighton, MA 02135, USA}
\author{Anshuman Singh}
\affiliation{Raytheon BBN Technologies, Cambridge, MA 02138, USA}
\author{James Leatham}
\affiliation{Raytheon, El Segundo, California 90245, USA }
\author{Marko Lon\v{c}ar}
\affiliation{John A. Paulson School of Engineering and Applied Sciences, Harvard University, Cambridge, MA 01238, USA}
\author{Moe Soltani}
\affiliation{Raytheon BBN Technologies, Cambridge, MA 02138, USA}

\begin{abstract}
Solving large-scale computationally hard optimization problems using existing computers has hit a bottleneck. A promising alternative approach uses physics-based phenomena to naturally solve optimization problems wherein the physical phenomena evolves to its minimum energy. In this regard, photonics devices have shown promise as alternative optimization architectures, benefiting from high-speed, high-bandwidth and parallelism in the optical domain. Among photonic devices, programmable spatial light modulators (SLMs) have shown promise in solving large scale Ising model problems to which many computationally hard problems can be mapped. Despite much progress, existing SLMs for solving the Ising model and similar problems suffer from slow update rates and physical bulkiness. Here, we show that using a compact silicon photonic integrated circuit optical phased array (PIC-OPA) we can simulate an XY Hamiltonian, a generalized form of Ising Hamiltonian, where spins can vary continuously. In this nanophotonic XY Hamiltonian solver, the spins are implemented using analog phase shifters in the optical phased array. The far field intensity pattern of the PIC-OPA represents an all-to-all coupled XY Hamiltonian energy and can be optimized with the tunable phase-shifters allowing us to solve an all-to-all coupled XY model. Our results show the utility of PIC-OPAs as compact, low power, and high-speed solvers for nondeterministic polynomial (NP)-hard problems. The scalability of the silicon PIC-OPA and its compatibility with monolithic integration with CMOS electronics further promises the realization of a powerful hybrid photonic/electronic non-Von Neumann compute engine. 
\end{abstract}

\maketitle

\section{Introduction}
NP-hard optimization problems are of interest to many fields including finance, cryptography, medicine, and biology, but  solutions to such problems cannot be guaranteed to be found in polynomial time, and traditional methods for solving such problems require resources which grow exponentially with problem size. In an effort to find efficient methods of solving, NP-hard combinatorial optimization problems can be mapped to Ising Hamiltonians \cite{nishimori, lucas, byrnes2022}, and continuous variable optimization problems can be mapped to the XY Hamiltonian \cite{kalinin2018}. High performance methods for solving XY and Ising Hamiltonians can benefit from non-traditional physics-based solvers which can harness properties unique to the physical systems of implementation to realize lower resource-consuming algorithms which cannot be implemented on conventional processors. Recently, promising works on such physics-based solvers have included quantum annealers \cite{Johnson2011, ross2022}, gate-based quantum computers \cite{qaoa}, trapped ions \cite{kimisingion}, optical parametric oscillators \cite{hamerlyCIM, marandi2014_parametric_oscillator, mcmahon2016_programmable, inagaki2016, strinati2021, ng2022, vadlamani2020, vadlamani2022, cim_honjo, latifpour2022, hyperspin}, stochastic nanomagnets \cite{sutton2017}, coupled lasers \cite{Latifpour2020, parto2019}, and spatial light modulators \cite{pierangeliPRL, pierangeliadiabatic, Fang2021, sun2022, gigan2022, kumar2020}.


\begin{figure*}[ht]
\centering
\includegraphics[width=17cm]{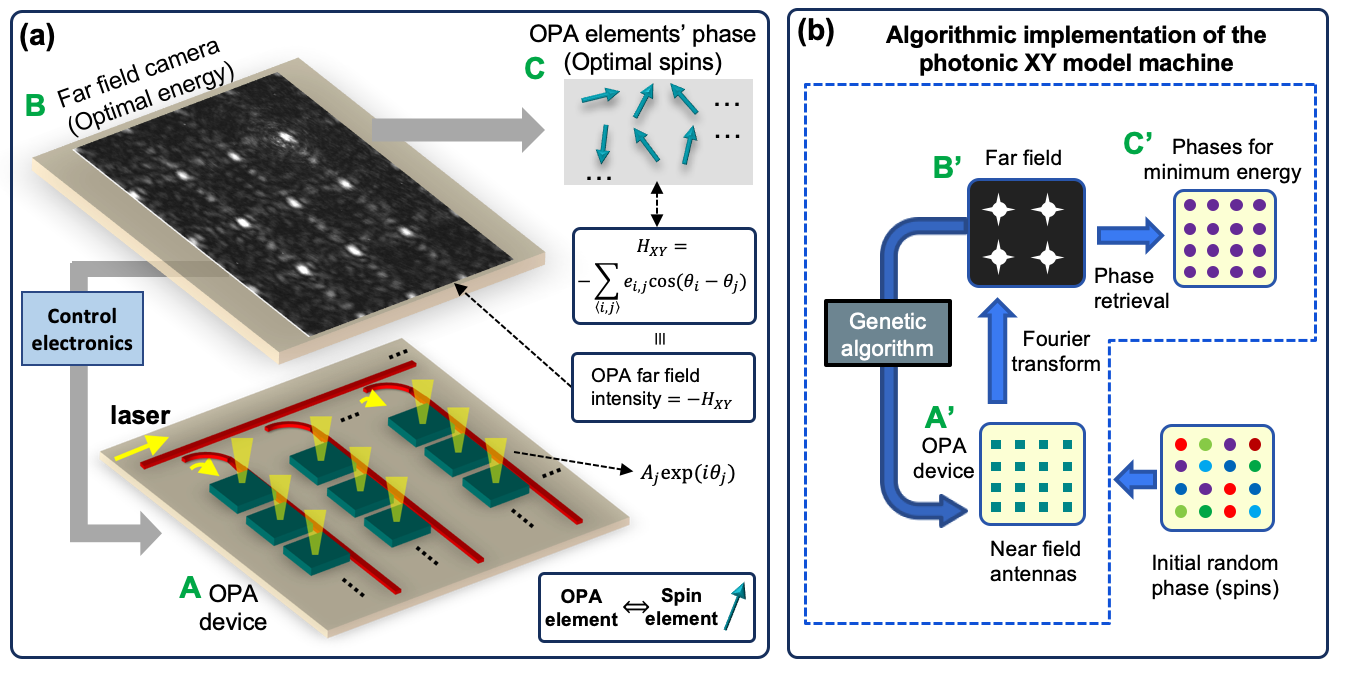}
\caption{
\textbf{General view of the photonic hardware used for the XY Hamiltonian solver, and the algorithmic approach for the solver}. (a) A diagram showing the XY model solver implemented with a PIC OPA. The phases of the optical phased array are encoded in the light emitted from each emitter pixel (A, A'), and controlled via voltage applied to the resonator phase shifters. In the focal plane (B, B'), the far field pattern is imaged and XY Hamiltonian energy is calculated, then voltages applied to each resonator phase shifter are modified over each optimization iteration to minimize or maximize energy through a feedback process. Phases are extracted using a phase-retrieval algorithm (C, C'). 
(b) A schematic showing the process of solving the XY model with a PIC OPA. Voltage is applied to the device to tune phases on the antennas (A, A'), which in the far field (B, B') forms a pattern from which the XY energy is calculated and used to feed back to tune the voltages. After the optimization process completes, the phases are extracted from the lowest energy far field pattern (C, C').  }
\label{fig:setup}
\end{figure*}

Photonic Ising solvers using programmable spatial light modulators (SLMs) can have considerable advantages over other Ising solvers due to the massive parallelism of spatial optical modes with no modal crosstalk in free-space. SLMs encode Ising spins in the phase of light and can simulate many spin all-to-all coupled Ising models \cite{pierangeliPRL}, as well as XY models with arbitrary coupling strengths \cite{ouyang2024}. In an SLM-based Ising solver, the propagation of light from the near field (directly emitted from the device) to the far field (where the camera or photodetector is placed) performs the Ising energy calculation, and the speed of such calculation is only limited by the detector and the backend electronics. Despite these numerous advantages, these conventional SLMs which are liquid-crystal based can be physically large, bulky, and slow. Significantly, current top-of-the-line liquid crystal SLMs have refresh rates two or more orders of magnitude lower than the phase-shifter modulation rates for on-chip devices \cite{gigan2022, aflounti2015, vcsel2023}. The limitations of traditional SLMs and increasing progress in development of on-chip SLM devices \cite{sun2013, poulton2019electro, xie2019IIIV} provide motivation to look for faster and more compact photonic approaches for solving NP-hard problems. In addition, the application of such photonic processors in solving XY model problems remains relatively unexplored, despite the existence of algorithms for efficient XY model optimization in gain-dissipative classical and quantum systems \cite{berloff2017}.

In this paper, we employ an on-chip and low power silicon photonic integrated circuit optical phased array (PIC-OPA) \cite{mvc_phasedarray2023} as a high-speed programmable SLM to solve for the energy minimum and corresponding spin configurations of an all-to-all coupled XY model. The XY model generalizes the Ising model by allowing spin to vary continuously, and can be used to solve optimization problems with continuous variables. In the all-to-all coupled XY model implemented by our PIC-OPA, every spin is coupled to every other spin with couplings that are a function of the OPA device geometry and spin location. Through implementation with silicon nanophotonic technology, our PIC-OPA XY model solver benefits in particular from fast spin refresh rates of 300 kHz and low power consumption of the analog phase shifters \cite{mvc_phasedarray2023}. Using analog phase shifters, the PIC-OPA processor naturally allows the simulation of a system with continuous XY model spins. 

Though more general XY Hamiltonians with varying connectivity and spins can be mapped to silicon PIC-OPAs \cite{luo_multiplexing}, the OPA circuit in this work only programs the spins, i.e. the phase shifters, resulting in an all-to-all XY model with equal connectivity between the graph edges. In a future work the OPA circuit can be redesigned to have both phase and amplitude control per antenna to enable programming of the spins as well as coupling strength between the spins. Using our OPA architecture, we find the energy minimum of an XY Hamiltonian and solve for the corresponding minimum spin configuration in the Fourier domain through a phase retrieval process. Our work demonstrates the potential of OPAs as compact, efficient and scalable solvers of NP-hard problems.

\section{Experiment}

In this work, we employ an 8x8 optical phased array OPA architecture with compact overcoupled ring resonator phase shifters \cite{mvc_phasedarray2023, hugo2019} to experimentally implement and solve a 64-node all-to-all coupled XY model. Photonic emission from the OPA imaged in the far field provides a physics-based calculation of XY model energies for our experiment.

An OPA can be used to encode and solve Ising and XY models. For a general Ising model, the interaction portion of the Ising Hamiltonian $H$ is equal to 
\begin{align}
\label{eq:isingeqn}
    H = - \sum_{<i,j>} e_{ij} Z_i Z_j
\end{align}
where $<i, j>$ denote all nodes $i$ and $j$ connected by an edge and $e_{ij}$ are edge weights. An OPA with an array of antennas each with independent phases constrained to equal either $0$ or $\pi$ can be mapped to an array of Ising spins \cite{pierangeliPRL}. The interference pattern produced from light emitted by an OPA, when summed and normalized, equals the energy of the Ising Hamiltonian. The interference terms between each antenna or spin give the couplings or edge weights $e_{ij}$.

In addition to Ising model problems, OPAs can also be used to encode and solve XY model problems. The XY model is the continuous phase counterpart of the Ising model, and the spin-spin interaction portion of the XY Hamiltonian is equal to \cite{berloff2017} 
\begin{align}
\label{eq:xyeqn}
    H =  - \sum_{<i,j>} e_{ij} \text{cos} \left( \theta_i - \theta_j \right)
\end{align}
where $<i, j>$ denote all nodes $i$ and $j$ connected by an edge, $e_{ij}$ are edge weights, and $\theta$ is constrained to equal any value between $[-\pi, \pi]$ \cite{berloff2017}. As in the Ising model case, the summed pixel intensity of the far field interference pattern emitted from an OPA is equal to the negative magnitude of the energy of the XY Hamiltonian. However, because the phases of the OPA antennas are no longer confined to be either $0$ or $\pi$ as in the Ising model, but instead can take any value between $[-\pi, \pi]$, XY model problems have a larger solution search space but also may have a greater density of near-solutions near the minimum. Table \ref{tab:mytable} shows a summary of a comparison of spin lattice systems and our phased array system, in the context of solving XY problems.

\begin{figure*}[ht]
\centering
\includegraphics[width=16cm]{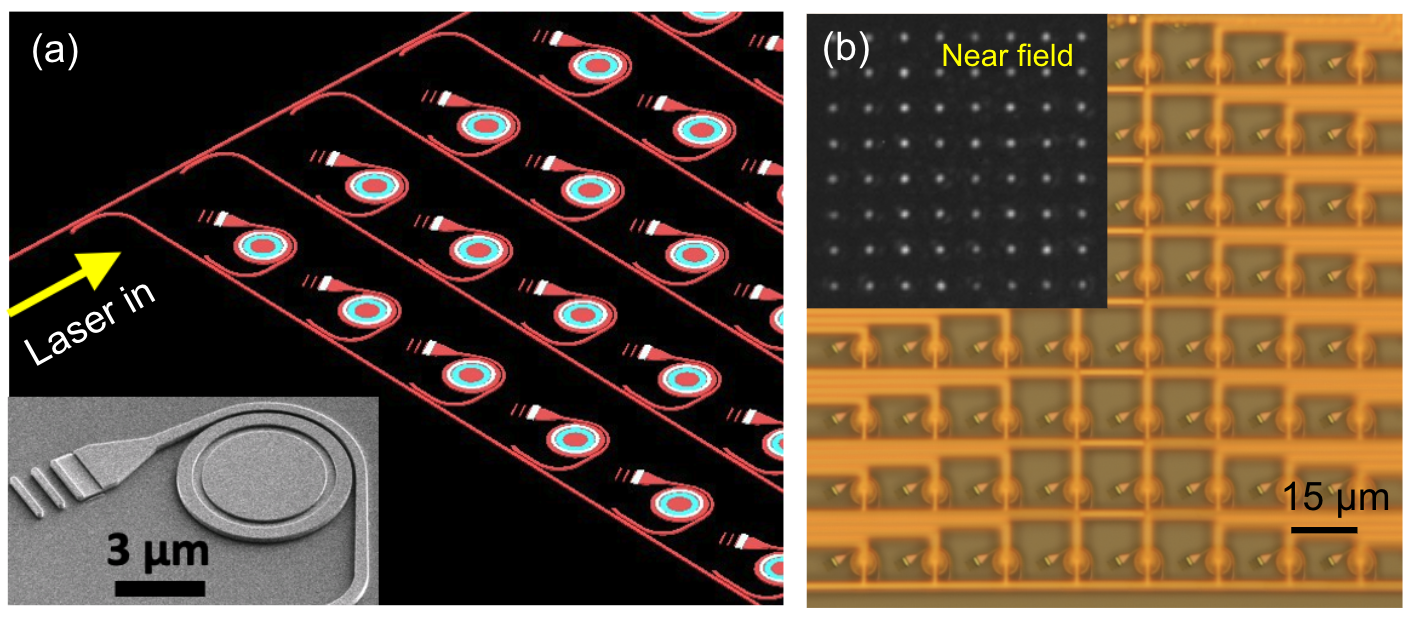}
\caption{
\textbf{Detailed structure of the PIC-OPA device used for the XY Hamiltonian solver.} (a) A 3D rendering of a portion of the 8x8 PIC-OPA chip, with a yellow arrow showing where the laser light enters before branching to reach each of the phased array elements and antennas. The inset shows a scanning electron microscope image of one phased array element consisting an overcoupled ring resonator phase shifter and an antenna. The electronic control of the phase shifter is not shown in this image. (b) A microscope image of the 8 by 8 optical phased array device. An inset shows a camera image of the emission from each antenna, taken in the near field. }
\label{fig:device}
\end{figure*}

\begin{table}[]
\caption{\label{tab:mytable} A summary of the key differences between the XY model with spin lattices, and the XY model implemented with an optical phased array.}
\begin{ruledtabular}
\begin{tabular}{l l l}
                                                                                  & Spin lattice                                                                              & Phased array                                                                \\ \hline
\multicolumn{1}{l}{Spin}                                                        & $Z_j $                                                                               & $ e^{i \theta_j}$                                              \\ \hline
\multicolumn{1}{l}{\begin{tabular}[c]{@{}l@{}}Description \end{tabular}} & {\begin{tabular}[c]{@{}l@{}}Lattice of spins \\ with  continuously \\ varying spin values \end{tabular}} & \begin{tabular}[c]{@{}l@{}}Lattice of \\ phased array elements \\ with continuously \\ varying phase\end{tabular} \\ \hline
\multicolumn{1}{l}{\begin{tabular}[c]{@{}l@{}}Energy\\ expression\end{tabular}} & XY Hamiltonian                                                                            & \begin{tabular}[c]{@{}l@{}}Phased array far field \\ intensity\end{tabular} \\ 
\end{tabular}
\end{ruledtabular}
\end{table}

 To solve the XY model using the OPA, first laser light with a wavelength near the overcoupled resonance of the ring resonators (here $\sim 1510$ nm) is routed into the OPA. The light is emitted through the antennas on-chip, and the resulting light emission pattern in the far field is imaged. To find the energy associated with a given configuration of spins (correspondingly, configuration of phases on each of the antennas), the total pixel intensity over an integer number of periods of the far field pattern is summed. After energy is calculated for a given configuration of spins, a genetic algorithm is used to optimize the voltages applied to each ring resonator to increase far field intensity and to solve for the XY model minimum. No information from the target Hamiltonian is required or used in the process of solving. Fig. \ref{fig:setup} shows a schematic of the process for solving the XY model, including a high-level view of the OPA device.

Once the XY model minimum energy is found through the genetic algorithm optimization process, the corresponding lowest energy spin configuration is found by retrieving the phase of each of the antennas in the OPA. In principle phase estimation at each antenna would be possible through a calculation of voltage applied to phase shift expected given previously characterized voltage to phase shift response curves. However, in practice fabrication inhomogeneities between ring resonators, thermal crosstalk between heaters, and thermal drift make such a calculation infeasible. Instead, we retrieve the XY model phases using the Gerchberg-Saxton phase retrieval algorithm \cite{gs, Fienup82}. 

Fig. \ref{fig:device}a shows a 3D rendering of a portion of the OPA device with 8x8 elements  used for the experiments. In this device, laser light enters the OPA through a waveguide, and through a tap coupler is split into eight horizontal bus waveguides with equal power. For each horizontal bus waveguide there are eight tap couplers to extract the light and pass to a ring resonator phase shifter before feeding into an antenna. All tap couplers are designed such that the antenna elements receive equal intensities. With appropriate resonator and waveguide coupling design we achieve strong overcoupling for the resonator phase shifters. A unit cell of this OPA has dimensions 15 $\mu m$ x 15 $\mu m$, and includes a resonator phase shifter with a radius of 2.75 $\mu m$, a grating antenna, and the tap coupler for inputted light. The use of an overcoupled ring resonator as a phase shifter provides fast and low power phase tuning. More details of the design and fabrication of this devices can be found in our prior work \cite{mvc_phasedarray2023}. Fig. \ref{fig:device}b shows the near field optical emission from the device, as well as a microscope image of the device, showing eight rows and eight columns of overcoupled ring resonator phase shifters with antennas. 
The overcoupled ring resonator acts as a low power phase shifter for light which couples from a waveguide to the ring resonator before emitting from an antenna. The phases of light emitted from each antenna, $\phi_j$ each encode an XY model spin. Phase at each antenna is controlled by adjusting the voltage applied to a doped resistive heater on each ring resonator.


\begin{figure}[ht]
\centering
\includegraphics[width=8cm]{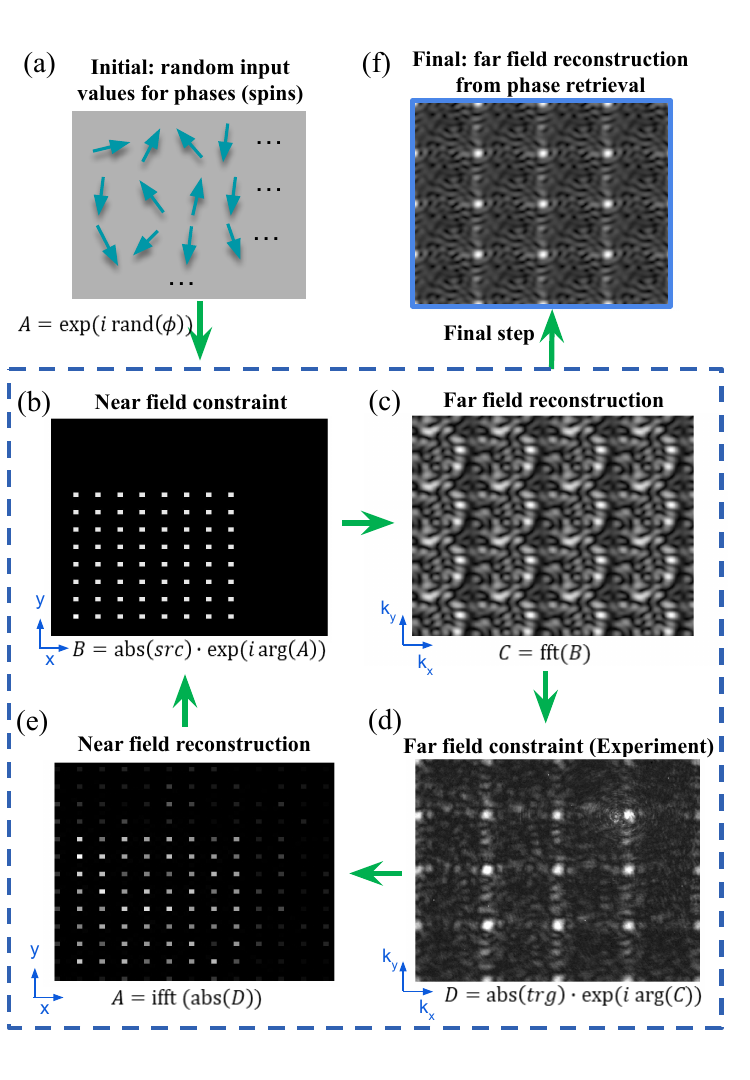}
\caption{\textbf{A schematic showing the phase recovery from the experimental data using the Gerchberg-Saxton algorithm.} First, (a) an array of random phases are generated. Then, (b) a near field constraint (\texttt{src}) according to the array geometry is multiplied by the random phases. Next, (c) the Fourier transform of the image is taken, and following that, (d) the far field constraint (experimental data or \texttt{trg}) is imposed, and (e) an inverse Fourier transform performed. The images in panels (b), (c), (d), and (e) show the results of the first iteration in this example. These steps are repeated over many iterations until the image after step C resembles the image after step D. In the example shown, (f) shows the reconstructed far field C after 10,000 iterations. }
\label{fig:gerchberg-saxton}
\end{figure}


\section{Results}

The XY model problem addressed in this work is that of an all-to-all coupled graph with 64 nodes. Though a uniform OPA with equivalent emission intensity from each antenna corresponds to an all-to-all coupled XY model, the edge weights in the model are modified by a constant phase dependent on the relative spatial distances between each on-chip antenna. Given a uniform periodic OPA with uniform emission intensities from each antenna, the weight products $w'_i w'_j$ for two antennas, antenna $i$ and antenna $j$, with the corrections for the finite distances between antenna are given by (see Appendix \ref{appendixa})
\begin{align}
\label{eq:weights}
    w'_i w'_j = {4 w_i w_j L_x L_y}  \: \text{sinc} ( \alpha \Delta m_{ij} L_x ) \: \text{sinc} ( \alpha \Delta n_{ij} L_y ),  
\end{align}
where $\alpha \equiv 2 \pi d/(\lambda_0 z_0)$, $L_x$ and $L_y$ are the horizontal and vertical distances over which the pixels in the far field pattern are summed, and $m_i$, $m_j$ ($n_i$, $n_j$) are the row and column number of the phased array element $m$ ($n$), such that $\Delta n_{ij} = (n_i - n_j)$ and $\Delta m_{ij} = (m_i - m_j)$. When $L_x$ and $L_y$ are constrained to be equal to an integer number of periods in the far field, the dependence on distance over which far field is integrated disappears. Eq. \ref{eq:weights} is valid near the center of the far field interference pattern where diffraction effects from the finite size of the antennas do not dominate, and experimental data is taken from this center region.

Using the 8x8 uniform OPA described above, we solve for the minimum energy and phase configurations for the corresponding XY model Hamiltonian. A schematic showing the process of phase retrieval using the Gerchberg-Saxton algorithm is shown in Fig. \ref{fig:gerchberg-saxton}. We use the constraint of a uniform array structure in the near field and experimental images of the far field interference pattern as the far field constraint.

In order to decrease the impact of noise, defects in periodicity, and non-uniformities in the far field image on the phase retrieval result, the far field image area used for phase extraction encompasses four periods of the diffraction pattern in the far field. The choice of four periods was determined through running optimization tests which employed the genetic algorithm without the Gerchberg-Saxton algorithm in order to optimize the far field to a specific sample pattern. Using too few far field periods can result in a retrieved phase map which is not representative of the actual phases due to failing to account for small but significant non-periodicity and imperfections between each far field unit cell. Using more far field periods implements an averaging over noise effects and more accurate phase retrieval. However, the camera and setup, as well as the need to maintain a high resolution image for each of the individual unit cells in the far field, set a constraint on the maximum number of periods that can be used.

\begin{figure}[ht]
\centering
\includegraphics[width=9cm]{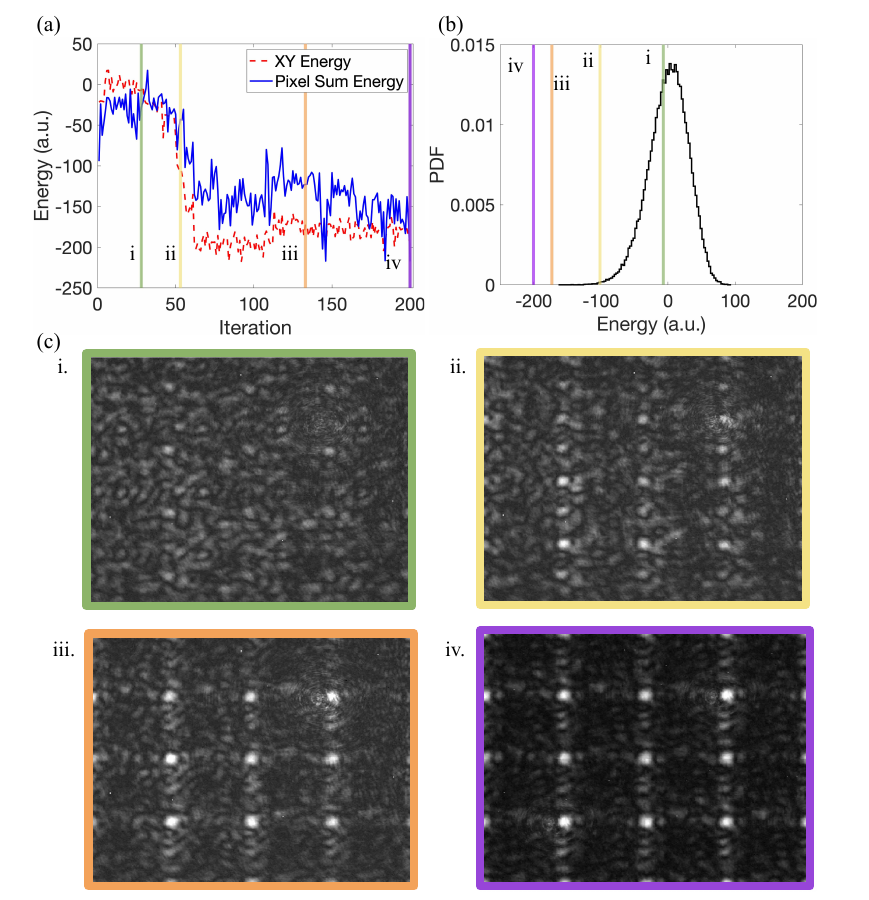}
\caption{\textbf{The energy minimization results for the OPA XY model solver for one optimization run.} (a) XY model energy and summed image pixel intensity for each iteration of an optimization for an energy minimum. (b) The simulated energy probability density function for the regular 8x8 OPA, with lines overlaid showing the experimental energies corresponding to the energies marked in (a) shown in different colors. (c) Experimental data showing the far field images corresponding to the energies marked in matching colors in (a) and (b). }
\label{fig:xyenergy}
\end{figure}

Fig. \ref{fig:xyenergy}a shows the energy minimization results for the 64 node all-to-all coupled XY model. The energies calculated through summing pixel intensity follow the energies calculated from the XY model Hamiltonian after phase extraction, signifying successful phase retrieval. As seen in Fig. \ref{fig:xyenergy}a, the variation of the energy per iteration decreases non-monotonically and is somewhat noisy. This noise in the summed pixel energy arises due to a large ratio of baseline image intensity to signal intensity. In addition, part of this noise as well as the non-monotonic behavior of the energy plot arises inherently due to the genetic algorithm process.  Appendix B contains further details on the sources of noise in this experiment.

Fig. \ref{fig:xyenergy}b shows the minimum energy found using the OPA (in purple), as well as other energies at earlier points during the optimization process (shown in green, yellow, and orange) overlaid on the simulated energy probability density function. As shown in Fig. \ref{fig:xyenergy}b, the final energy recovered is a low energy or near minimum solution to the XY model Hamiltonian. Fig. \ref{fig:xyenergy}c shows the experimental images at each of the corresponding points throughout the optimization process, including (purple boxing) the minimum. As visible in the figure, the experimentally found energy minimum is at the tail of the probability density function.

\begin{figure}[ht]
\centering
\includegraphics[width=9cm]{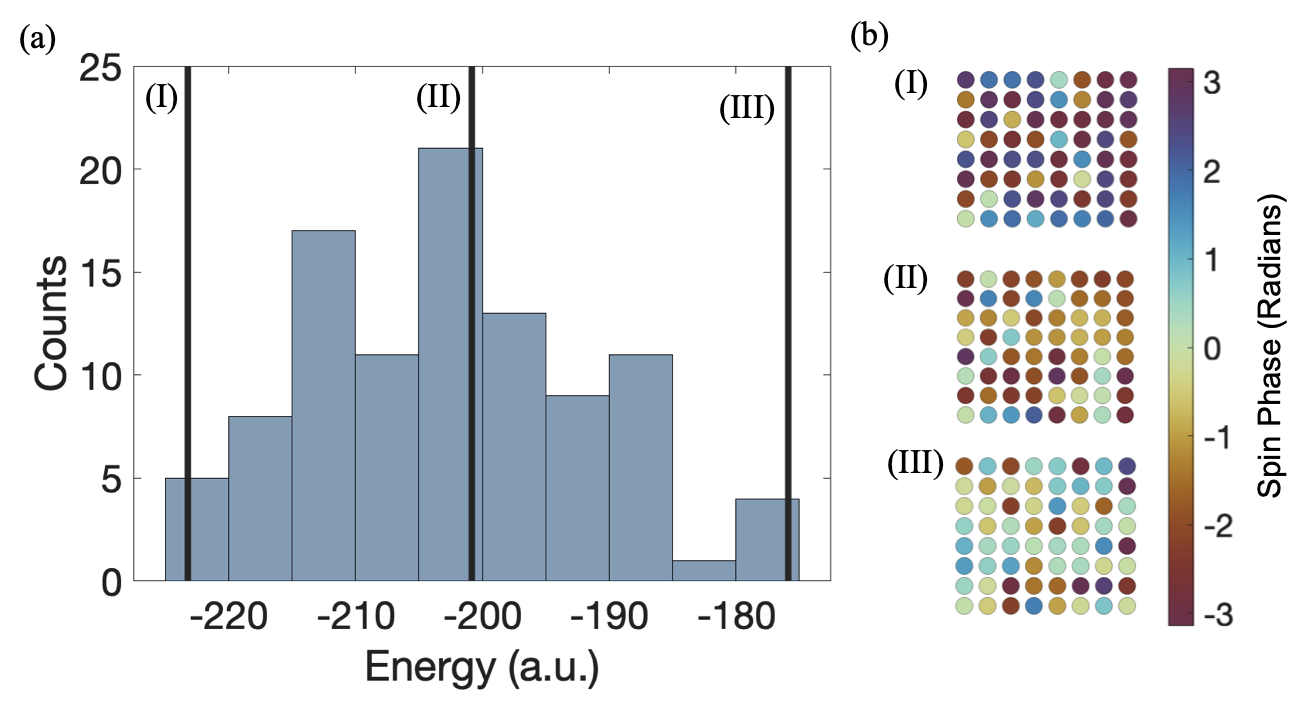}
\caption{\textbf{Dependence of XY model energy (calculated from retrieved phases) on the random seed in the Gerchberg-Saxton algorithm.} (a) Distributions of solved XY model energies for differently seeded phase retrievals, with selected retrieved phase sets shown as insets marked by Roman numerals. For 100 random seeds, a 10,000 iteration Gerchberg-Saxton phase retrieval is performed using far field data and the histogram of the resulting energies are plotted. (b) Three phase configurations associated with different random seeds. Their corresponding energies are plotted as vertical lines (a) and labeled.  }
\label{fig:gshistogram}
\end{figure}

Far field image noise results in a distribution of retrieved phases and distribution of energies given different initial random seedings of the Gerchberg-Saxton algorithm. The far field reconstruction from phase retrieval (see Fig. \ref{fig:gerchberg-saxton}) can never perfectly match the far field constraint of the experimental data due to optical scattering and other aberrations in the far field image data. Fig. \ref{fig:gshistogram}a shows the histogram of solved XY model energies given different Gerchberg-Saxton algorithm random seeds at the final phase extraction step. Though the energy spread is clustered at the tail of the simulated probability density function (see Fig. \ref{fig:xyenergy}), there is a spread in energy which is dependent on the number of Gerchberg-Saxton iterations. Three examples of solved phase configurations are shown in Fig. \ref{fig:gshistogram}b, with their corresponding energies marked in the histogram. As their corresponding energies decrease, the phases become more uniform within each distribution.

\section{Discussion}
The OPA-based XY Hamiltonian solver calculates the XY model Hamiltonian energies purely photonically, and uses a genetic algorithm to minimize the energy.  Our approach can be compared to a solver which combines a genetic algorithm for function minimization with a computer calculation to find the Hamiltonian energy. The energy calculation portion of our approach has the benefit of a constant time scaling with Hamiltonian size. We can also leverage heuristic recurrent optimization algorithms \cite{carmes2020} into our solver.

Our XY model solver provides the benefit of fast multiplication of spin-spin coupling via free-space photonic propagation and interference to generate the far field, as well as the benefit of efficient phase-retrieval for OPAs. The composite nature of our XY model solver enables further fine-tuning or even replacement of the constituent genetic algorithm or Gerchberg-Saxton phase retrieval algorithm. As an example, one can replace the genetic algorithm with a Monte Carlo or gradient-based search algorithm \cite{sun2022, pierangeliPRL}. In simulated tests, the genetic algorithm showed the fastest and most efficent convergence when compared to other optimization algorithms, including Matlab's fminsearch and particle swarm algorithms (see Appendix C for details).

In our XY model solver algorithm, the phase retrieval process is a time limiting step, as it requires thousands of Fourier and inverse Fourier transforms on the experimental far field image. The retrieved phases and corresponding energies for the solved XY model are dependent on the random seed in the Gerchberg-Saxton phase retrieval step with a variance in energies determined by the number of Gerchberg-Saxton iterations. However, the phase retrieval step is performed post-solving, without augmenting the solving process, which uses the summed pixel intensity of the far field image for the Hamiltonian energy. Significantly, the Gerchberg-Saxton algorithm scales with the number of pixels in the far field image, rather than scaling directly with the size of the OPA array.


Compared to an alternative phase readout method of converting voltages applied to the heaters to phases through a characterization of phase shifter response with voltage, the Gerchberg-Saxton phase retrieval technique has the advantages of lacking dependence on device calibration and is a more direct measurement of the phases or XY model spins. The Gerchberg-Saxton technique is also more practical and feasible than a phase readout method based on the characterization of individual phase shifter response with voltage. The characterization based on the latter approach would be primarily difficult due to the many measurements required to individually characterize each phase shifter especially for large size arrays, as well as to account for crosstalk.

The characterization of the response of the phase shifters with applied voltage would require at least 64 measurements for an array of 64 phase shifters. However, the magnitude of the response of each resonance with change in voltage applied is nonlinear and proportional to the square of the voltage, and the curves modeling the wavelength shift of resonance as voltage is applied to a resonator’s heater vary slightly but significantly with each resonator. Measurements at several voltages will need to be performed for each phase shifter.

The complexity of the characterization process increases when one considers thermal crosstalk between phase shifters. To account for this crosstalk, at first approximation, for all heaters $i = [1.. 64]$ and resonators $j = [1..64]$, one would need to measure the resonance shifts produced by heater i acting on resonator j to produce a 2D matrix, which would be of dimension 64 x 64 to account for the effect of each resonator on each resonator. This approximation likely will not be sufficient, however. A better approximation should account for the effects of two or more heaters on a third resonator, because the heating effects are nonlinear and cannot simply be added when finding the phase-shifting effect of multiple simultaneous heaters on a given resonator. This will drastically increase the size of the required matrix. To perform a careful calibration, one will also then wish to repeat the matrix measurement for different voltages, which further scales the problem to become impractical. The Gerchberg-Saxton approach circumvents all of these issues by directly reading out the phase without attempting to determine it from the applied voltages.

In this work we only vary the phase of the OPA phase shifters which represent the spin in the XY model, while the spin-spin coupling weights are constant. However, the PIC-OPA circuit design can be updated to allow for varying the spin-spin coupling weights. An OPA with independent intensity control in each antenna will allow for the solving of XY model problems for graphs with differing edge weight pairs (see the simulations of such a case in Appendix D). To implement an OPA with both phase and amplitude control for modulating both spin and spin-spin coupling, we can include an on-chip optical attenuator before each phase shifter. An optical attenuator can be implemented either using a waveguide-based charge-injection attenuator by integrating a PN junction with the silicon waveguide before the phase shifter, or using another microresonator in the path which is undercoupled and can induce attenuation with minimal phase perturbation. In this latter scheme, we have one strongly overcoupled resonator for the phase shifter purpose, and one undercoupled resonator for the attenuation purpose.

Though in this work we used thermo-optic based phase shifters that induce thermal crosstalk, one can reduce the crosstalk by increasing the pitch of the OPA to increase the phase shifter spacing. Alternatively in future OPA designs one can explore other phase shifter mechanisms such as electro-optic approaches that have negligible to no crosstalk. 

\section{Conclusion}
In conclusion, we have implemented and solved an all-to-all coupled XY model Hamiltonian using a silicon nanophotonic two-dimensional optical phased array made of 64 elements. For a uniform optical phased array, we found energies which populate the tail of the XY model energy distribution. We calculated the XY model energy in two ways: through summing the pixel intensity of the far field image, and through retrieval of the phases on each antenna in the OPA. The two energies matched each other for each optimization iteration, up to the noise from image intensity fluctuations.

Though in this work we used a 64-element OPA, the OPA circuit can be scaled to a much larger number of elements offering the ability to solve XY Hamiltonian problems with larger graphs. The current optimization process uses a genetic algorithm, and the final phase (spin) extraction uses the Gerchberg-Saxton algorithm. The Gerchberg-Saxton algorithm is quite efficient and almost independent of the array size. A future work can explore alternative algorithms to the genetic algorithm for more efficient computation.

As the OPA size scales up, use of a camera with high resolution during the optimization and phase retrieval process can help to increase the image detail and information retrieved from each period of the far field image. In addition, the use of a high-speed camera will enable faster image data transfer to the main processor enabling a faster XY Hamiltonian solver. The OPA device used in this work only programs the spin values and not the spin-spin couplings, limiting the XY Hamiltonian to a modeling a single uniform all-to-all connectivity graph. However, spin-spin coupling reconfigurability can be implemented in next generation devices by introducing on-chip optical tunable attenuator to control the intensity from each antenna independently. 
As a proof of concept, this work also shows promise for other application-specific integrated circuit computing with on-chip OPAs, including machine learning \cite{pierangeliPEL} and adiabatic annealing \cite{pierangeliadiabatic}.

\section*{Acknowledgements}
MC acknowledges support from the Department of Defense (DoD) through the National Defense Science and Engineering Graduate (NDSEG) Fellowship Program. The authors thank Dr. Srikrishna Vadlamani from Massachusetts Institute of Technology for helpful discussions. 

This document does not contain technology or technical data controlled under either the U.S. International Traffic in Arms Regulations or the U.S. Export Administration Regulations.

\section*{Authors Declaration}
\section*{Conflict of Interest}
Moe Soltani and James Leatham have a patent on photonic integrated circuit phased array compute engine.

\section*{Data Availability Statement}
The data that support the findings of this study are available from the corresponding author upon reasonable request.

\appendix
\section{Derivation of uniform phased array XY model weights}
\label{appendixa}

In this appendix we derive Eq. \ref{eq:weights_appendix} for the optical phased array XY model Hamiltonian weights. Given a uniform periodic OPA, we find the weights $w'_i$ and $w'_j$ for two antennas, antenna $i$ and antenna $j$, corrected for the finite distances between each antenna in a uniform array of antennas. 

The equation for the array factor for the interference pattern for a two-dimensional uniform phased array with $N$ elements is in general
\begin{align}
   \text{AF}(x, y) =   \sum^N_{j=1} w_j e^{ i\left( \frac{2 \pi d}{ \lambda_0 z_0} \left( m_j \cdot x + n_j \cdot y \right) \right) + i \phi_{mn}^j}, 
\end{align}
where $m_j$, $n_j$, are the column and row number of the phased array element $j$, $\phi_{mn}^j$ is the phase on the $j$th array element at column $m$ and row $n$, $w_j$ is the intensity at antenna $j$, $d$ is equal to the spacing between the antennas, $\lambda_0$ is the wavelength of laser light, and $z_0$ is the distance between the imaging plane and the source.

Given $I$ is the the total summed pixel intensity (disregarding the finite size of the antennas, and only including array factor effects) for an image containing an integer number of periods of the interference pattern in the far field, and $C$ is a constant equal to the sum of each of the non-corrected weights, squared ($C = \sum_k^{64} |w_k|^2$), we can write
\begin{widetext}
\begin{align}
    H = -(I - C) = - \sum_x^\text{pixels} \sum_y^\text{pixels} & | \text{AF}| ^2 - C = - \sum_x^\text{pixels} \sum_y^\text{pixels} \sum_i^{64} \sum_j^{64}  w_i w_j e^{i \alpha \left( (m_i x + n_i y) - (m_j x + n_j y)\right)  + i (\phi_{mn}^i - \phi_{mn}^j) } + c.c. \\
    = - \sum_x^\text{pixels} \sum_y^\text{pixels} & \sum_i^{64} \sum_j^{64}  w_i w_j 2 \cos \left( \alpha \left( (m_i x + n_i y) - (m_j x + n_j y)\right)  +  (\phi_{mn}^i - \phi_{mn}^j) \right) \\
     = - \sum_i^{64} \sum_j^{64} & w_i w_j 2[ \cos (\phi_{mn}^i - \phi_{mn}^j) \sum_x^\text{pixels} \sum_y^\text{pixels} \cos \left( \alpha (m_i - m_j)x + \alpha (n_i - n_j) y \right)  + \\ \nonumber
     &\sin (\phi_{mn}^i - \phi_{mn}^j) \sum_x^\text{pixels} \sum_y^\text{pixels} \sin \left( \alpha (m_i - m_j)x + \alpha (n_i - n_j) y \right) ], \nonumber
\end{align}
\end{widetext}
where  $\alpha \equiv 2 \pi d/(\lambda_0 z_0)$.

Considering a sum over many pixels (number of pixels $\gg$ 64), we approximate the sums over $x,y$ as integrals. We also integrate over an integer number of periods in $x$, from $-L_x$ to $L_x$, and an integer number of periods in $y$, from $-L_y$ to $L_y$. Also let $\Delta m_{ij} = (m_i - m_j) $,  $\Delta n_{ij} = (n_i - n_j) $, $\Delta \phi_{mn}^{ij} = (\phi_{mn}^i - \phi_{mn}^j)$. Then, 
\begin{align}
    I - C =& \\ \sum_i^{64} \sum_j^{64} & w_i w_j 2 [ \cos \Delta \phi_{mn}^{ij} \int_{-Lx}^{Lx} \int_{-Ly}^{Ly} \cos \left( \alpha \Delta m_{ij} x + \alpha \Delta n_{ij} y \right) dx dy \nonumber \\ 
     & + \sin \Delta \phi_{mn}^{ij} \int_{-Lx}^{Lx} \int_{-Ly}^{Ly} \sin \left( \alpha \Delta m_{ij} x + \alpha \Delta n_{ij} y \right) \; dx \; dy  ] \nonumber
\end{align}
The sine terms integrate to 0 considering we integrate over an integral number of periods, leaving: 
\begin{widetext}
\begin{align}
     I - C = \sum_i^{64} \sum_j^{64}  - 2 w_i w_j  \cos \Delta \phi_{mn}^{ij} \frac{1}{\alpha \Delta m_{ij}}  \frac{1}{\alpha \Delta n_{ij}} [& \cos (\alpha \Delta m_{ij} L_x + \alpha \Delta n_{ij} L_y) - \cos (\alpha \Delta m_{ij} L_x - \alpha \Delta n_{ij} L_y)  \\
     &- \cos (- \alpha \Delta m_{ij} L_x + \alpha \Delta n_{ij} L_y) + \cos (- \alpha \Delta m_{ij} L_x - \alpha \Delta n_{ij} L_y)] \nonumber \\
     = \sum_i^{64} \sum_j^{64}   2 w_i w_j  \cos \Delta \phi_{mn}^{ij} & \frac{4}{\alpha^2 \Delta m_{ij} \Delta n_{ij}}  \sin ( \alpha \Delta m_{ij} L_x ) \sin ( \alpha \Delta n_{ij} L_y )   
\end{align}
\end{widetext}
So we recover an XY model Hamiltonian in which
\begin{align}
    H = -(I - C) = - \sum_{<i,j>} w'_i w'_j \cos \Delta \phi_{mn}^{ij}, 
\end{align}
where
\begin{align}
    w'_i w'_j = \frac{4 w_i w_j}{\alpha^2 \Delta m_{ij} \Delta n_{ij}}  \sin ( \alpha \Delta m_{ij} L_x ) \sin ( \alpha \Delta n_{ij} L_y ).   
\end{align}

The above expression can be also written as below

\begin{align}
\label{eq:weights_appendix}
    w'_i w'_j = {4 w_i w_j L_x L_y}  \: \text{sinc} ( \alpha \Delta m_{ij} L_x ) \: \text{sinc} ( \alpha \Delta n_{ij} L_y ),  
\end{align}

\section{Sources of noise in the XY Hamiltonian solver}
\label{appendixb}

\begin{figure*}[ht]
\centering
\includegraphics[width=17cm]{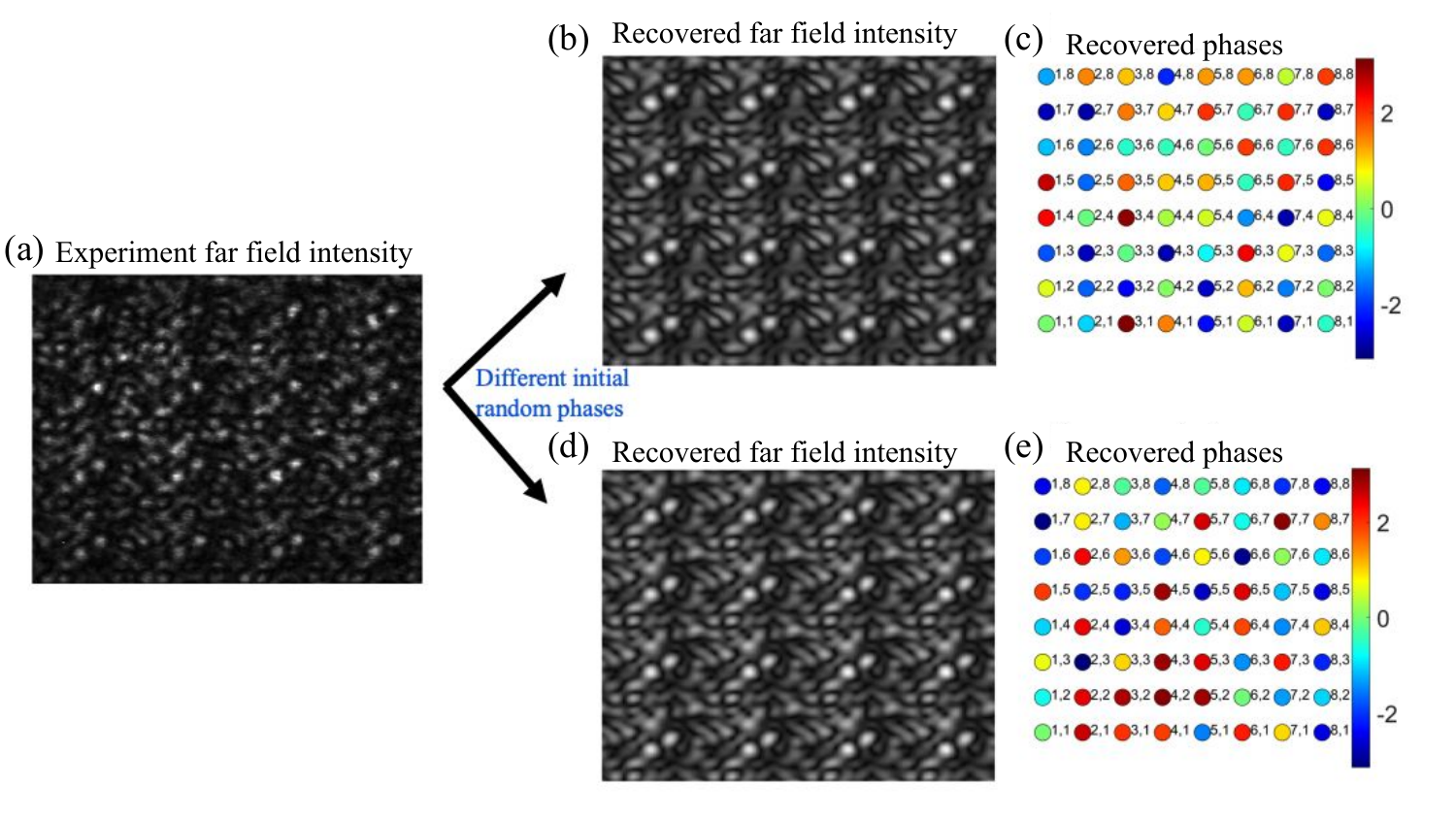}
\caption{ \textbf{Dependence of recovered far field and recovered phases on the random seed given the noisy experimental far field constraint in the Gerchberg-Saxton algorithm.} (a) An experimental image of four periods in the far field of our 8x8 PIC-OPA device before optimizing for the minimum of the XY Hamiltonian. When the phases are recovered from this image using the Gerchberg-Saxton algorithm, different phase maps will result which depend on the different random initial seeds used in the algorithm. (b) and (c) show respectively the resulting simulated recovered far field intensity and the corresponding phase for one random seeding of the Gerchberg-Saxton algorithm. (d) and (e) show respectively the resulting simulated recovered far field intensity and the corresponding phase for another random seeding which differ from the one used for (b) and (c).} 
\label{fig:differentseeds}
\end{figure*}

Various sources of noise are present throughout the process of solving the XY Hamiltonian using the OPA solver. One source of noise occurs due to fluctuations in image-wide scattering and intensity in the far field images used for the phase retrieval. Fluctuations in global image intensity can add noise to the energy calculated by summing the pixels in the far field image, creating or exacerbating non-monotonicities when plotted against optimization iterations.

Another source of noise in the retrieved phases stems from subtle non-identicalities such as the nonuniform scattered light in the far field unit cells, which prevent the far field reconstructed from the retrieved phases from ever exactly matching the far field constraint of the experimental data. Noise and scattering prevent each far field unit cell from being a perfect copy of one another. As a result, seeding the Gerchberg-Saxton algorithm with different random numbers during phase retrieval will result in different retrieved phases as illustrated in Fig. \ref{fig:differentseeds}.

The retrieved phases differ when seeded with different initial random phases. These sets of phases will not converge to match a true reference set even with an arbitrarily high number of iterations used in the Gerchberg-Saxton algorithm. However, the XY model Hamiltonian energies associated with these phases will be close in value.

Thermal crosstalk from the heated resonators may contribute towards delaying algorithm convergence and adding cycles to the genetic algorithm used to minimize the XY Hamiltonian. The largest wavelength shifts observed due to crosstalk in this experiment were 0.1-0.5 nm, which are large compared to the maximal wavelength shifts (1- 4 nm) observed to a resonator when the maximum voltage is applied directly.

The random nature of genetic algorithms means that during an optimization, the energy will not necessarily decrease monotonically with successive iterations. This can contribute to a jagged or noisy appearance of plots when energy is plotted against iteration number (see Fig. \ref{fig:xyenergy}a). In the genetic algorithm used in this work, the fitness of each individual in a population is evaluated according to a cost function. Here, an individual is a set of 64 voltages to apply to each of the 64 resonators. A fraction of the individuals with the lowest fitness are discarded from the population. The most fit individual is copied the necessary number of times to maintain the original population size. Then, the surviving individuals undergo combination, in which the individuals pair up to exchange a specified fraction of their 64 voltages, and mutation, in which a specified fraction of their 64 voltages are randomly modified to new voltage values. These steps, from the evaluation of individuals’ fitness to the combination and mutation, constitute an iteration.

Hyperparameters such as population size or fraction of population to undergo mutation are chosen for the genetic algorithm by running a series of parameter sweeps for two test reference far field patterns. Although in general genetic algorithms tend to lead to fitter individuals in the population after each iteration, this is not guaranteed due to the random nature of the algorithm, and especially when the parameters of the algorithm are tuned to permit greater randomness. However, particularly in highly multi-variable optimization problems, genetic algorithms remain powerful tools for accessing near-minimum and low energy solutions \cite{skaar98, katoch2021}.

\section{Simulated comparisons of optimization algorithms used for the OPA XY Hamiltonian solver}
\label{appendixd}

This work uses a genetic algorithm to minimize the XY Hamiltonian based on the image intensity of the far field. To determine the efficiency of the genetic algorithm compared to other algorithms, and to investigate the performance of this algorithm with increasing OPA array size, we perform the XY Hamiltonian optimization for simulated OPA far field data using the genetic algorithm and two other algorithms. The two algorithms we use for comparison are Matlab’s fminsearch algorithm and Matlab’s particle swarm algorithm.

\begin{figure}[ht]
\centering
\includegraphics[width=7cm]{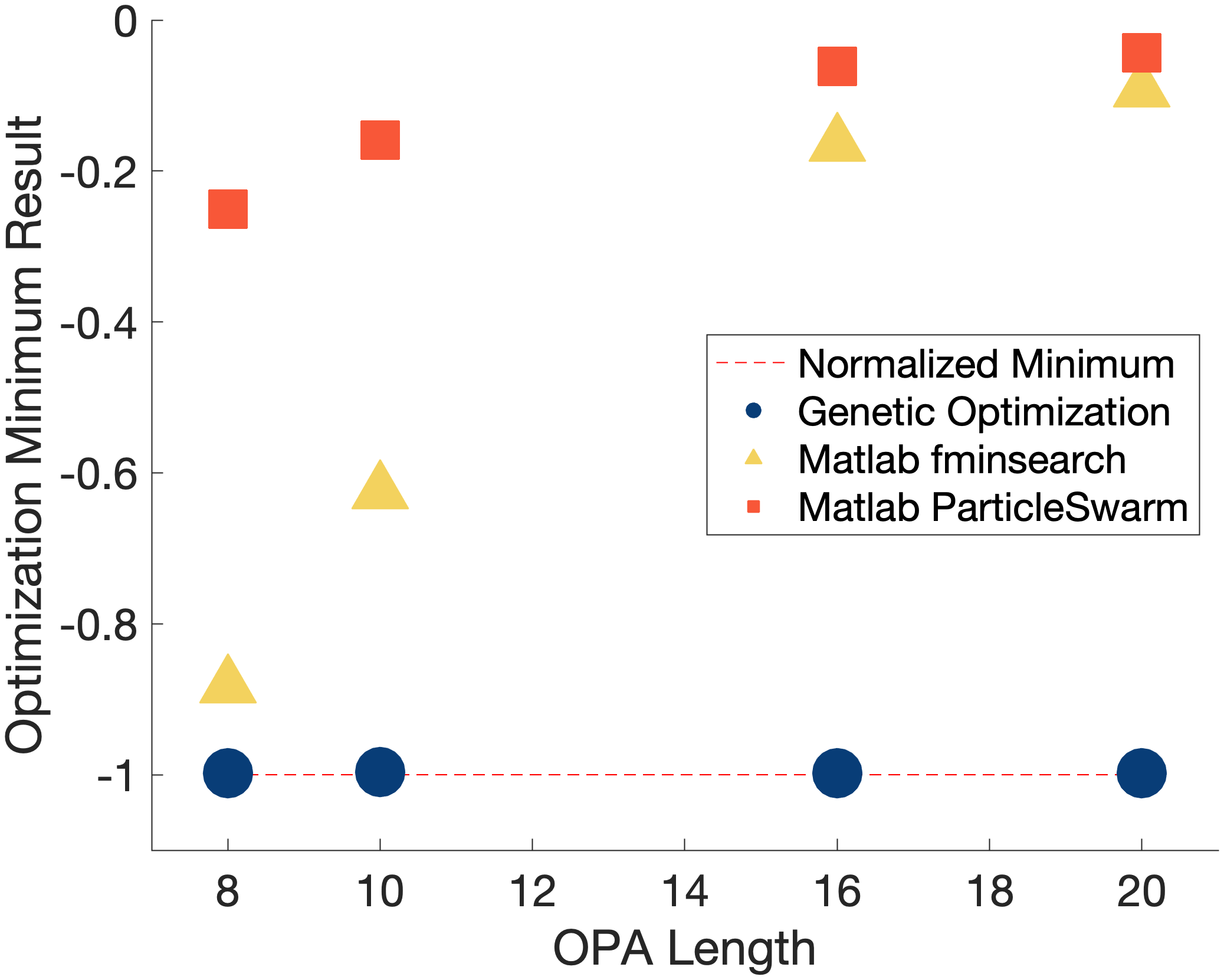}
\caption{\textbf{Simulated comparison of optimization algorithms for XY model Hamiltonian solving. } The minimum result returned from the simulation of optimization, normalized by the absolute  minimum of the equally weighted, all-to-all coupled XY model Hamiltonian, for square OPAs with different OPA lengths. Results are shown for different types of optimization algorithms including a genetic algorithm, Matlab’s fminsearch algorithm, and Matlab’s particle swarm algorithm.} 
\label{fig:simulatedoptimizations}
\end{figure}

Fig. \ref{fig:simulatedoptimizations} shows the minimums recovered by each optimization function for the equally weighted, all-to-all coupled XY model Hamiltonian represented by OPA with different lengths. The minimums plotted are normalized to the true minimum for each XY model problem. The data in this figure is produced from a simulation of our XY model Hamiltonian solver system (in which phases are controlled on a simulated OPA, which produces a simulated far field, and which the optimization function in question then minimizes the Hamiltonian simulated by the OPA). The figure shows the genetic algorithm closely approaches the minimum energy for each problem size, outperforming both fminsearch and the particle swarm algorithm. Based on the data from this figure, we conclude that the genetic algorithm is well suited to this particular optimization problem.

The energy efficiency can be estimated by comparing the number of function calls (calculations of XY Hamiltonian energy) for each optimization process. The genetic algorithm required only 3000 calls to simulate the far field in order to obtain a convergence close to the true minimum. The fminsearch algorithm and particle swarm algorithm required greater than 10,000 function calls to simulate the far field while failing to reach the same level of convergence.

\section{OPA XY Hamiltonian solver with antenna intensity and phase control for more general graphs}
\label{appendixc}

\begin{figure}[ht]
\centering
\includegraphics[width=8cm]{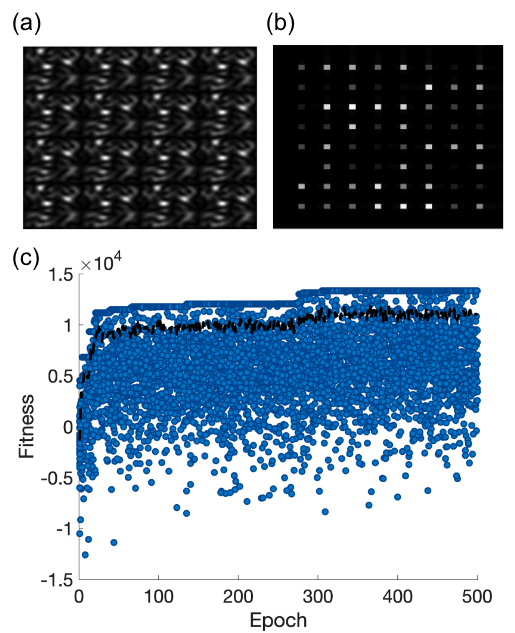}
\caption{\textbf{Simulations for an OPA XY model Hamiltonian solver with  phase and intensity control for each antenna.} (a) and (b) show respectively the simulated far field maximum solution and the corresponding near field image for a Mattis model graph maximization problem. For this simulation, the size of the OPA is 8x8 and each antenna is set to a different emission intensity. (c) The simulated fitnesses using a genetic algorithm to maximize the Mattis model Hamiltonian. The dashed black line indicates the average population fitness.}
\label{fig:generalmodel}
\end{figure}

As discussed earlier, our OPA design only allowed control of the antenna phase, and not the antenna emission intensity, limiting the XY Hamiltonian model solver to implementing and solving an all-to-all coupled graph of 64 nodes. In this section, we show that with the ability to control both antenna emission intensity and phase in next generation OPA devices, a broader category of Hamiltonians beyond all-to-all coupled Hamiltonians will be solvable. Though this described OPA providing both antenna emission intensity and phase control cannot solve XY models with arbitrary coupling schemes, it can be programmed to solve a class of XY models known as spin-glass Mattis models. In this appendix, we provide simulations to demonstrate the potential for the solvability of a larger class of graphs beyond constant edge weight all-to-all coupled graphs in next generation OPA devices.

Fig. \ref{fig:generalmodel} shows a simulated maximum energy problem for a spin-glass Mattis model. Mattis models\cite{pierangeliPRL} have coupling interactions (edge weights) $e_{ij}$ satisfying the proportionality $e_{ij} \propto \gamma_i \gamma_j$ for some parameters $\gamma_n$ specific to each antenna emitter $n$. Mattis models can be implemented in future generation OPAs through controlling the antenna emission intensity of the OPA. The intensity for each antenna $n$ will contribute to the term $\gamma_n$, and can be set to modify the XY model Hamiltonian which will be solved. The process for solving the Hamiltonian will follow the same method used for solving the constant weight all-to-all coupled model: a genetic algorithm optimizing on far field images from the OPA will provide feedback to evolve the OPA toward the XY model Hamiltonian solution, while the Gerchberg-Saxton method will provide phase retrieval on the far field image to retrieve the final solution. 

We present simulated results  to show the potential for solving XY model Hamiltonians with future generation OPA which have controllable antenna emission intensity. We simulate the resulting far field obtained by the simulated OPA corresponding to the maximum XY model Hamiltonian energy for a Mattis model with random weights chosen from a uniform distribution $[0, 1]$ (Fig. \ref{fig:generalmodel}a). We plot the corresponding simulated near field  of the OPA set to a configuration to model the described Mattis model (Fig. \ref{fig:generalmodel}b). Fig. \ref{fig:generalmodel}c shows the energies for each iteration in a simulated genetic algorithm, evolving the OPA (through changing the phases) towards the  solution, in this case, a maximum. These simulations show that with the added feature of antenna emission intensity control, OPAs can solve Mattis model Hamiltonians. In future work, adding wavelength multiplexing in addition to antenna emission intensity and phase control can allow arbitrary spin couplings to be implemented in the OPA, showing the promise of the platform for solving more general XY model Hamiltonians \cite{luo_multiplexing}.

\bibliography{sources}

\begin{thebibliography}{44}%
\makeatletter
\providecommand \@ifxundefined [1]{%
 \@ifx{#1\undefined}
}%
\providecommand \@ifnum [1]{%
 \ifnum #1\expandafter \@firstoftwo
 \else \expandafter \@secondoftwo
 \fi
}%
\providecommand \@ifx [1]{%
 \ifx #1\expandafter \@firstoftwo
 \else \expandafter \@secondoftwo
 \fi
}%
\providecommand \natexlab [1]{#1}%
\providecommand \enquote  [1]{``#1''}%
\providecommand \bibnamefont  [1]{#1}%
\providecommand \bibfnamefont [1]{#1}%
\providecommand \citenamefont [1]{#1}%
\providecommand \href@noop [0]{\@secondoftwo}%
\providecommand \href [0]{\begingroup \@sanitize@url \@href}%
\providecommand \@href[1]{\@@startlink{#1}\@@href}%
\providecommand \@@href[1]{\endgroup#1\@@endlink}%
\providecommand \@sanitize@url [0]{\catcode `\\12\catcode `\$12\catcode `\&12\catcode `\#12\catcode `\^12\catcode `\_12\catcode `\%12\relax}%
\providecommand \@@startlink[1]{}%
\providecommand \@@endlink[0]{}%
\providecommand \url  [0]{\begingroup\@sanitize@url \@url }%
\providecommand \@url [1]{\endgroup\@href {#1}{\urlprefix }}%
\providecommand \urlprefix  [0]{URL }%
\providecommand \Eprint [0]{\href }%
\providecommand \doibase [0]{http://dx.doi.org/}%
\providecommand \selectlanguage [0]{\@gobble}%
\providecommand \bibinfo  [0]{\@secondoftwo}%
\providecommand \bibfield  [0]{\@secondoftwo}%
\providecommand \translation [1]{[#1]}%
\providecommand \BibitemOpen [0]{}%
\providecommand \bibitemStop [0]{}%
\providecommand \bibitemNoStop [0]{.\EOS\space}%
\providecommand \EOS [0]{\spacefactor3000\relax}%
\providecommand \BibitemShut  [1]{\csname bibitem#1\endcsname}%
\let\auto@bib@innerbib\@empty
\bibitem [{\citenamefont {Nishimori}(2001)}]{nishimori}%
  \BibitemOpen
  \bibfield  {author} {\bibinfo {author} {\bibfnamefont {H.}~\bibnamefont {Nishimori}},\ }\href@noop {} {\emph {\bibinfo {title} {Statistical Physics of Spin Glasses and Information Processing: An Introduction}}}\ (\bibinfo  {publisher} {Oxford University Press},\ \bibinfo {year} {2001})\BibitemShut {NoStop}%
\bibitem [{\citenamefont {Lucas}(2014)}]{lucas}%
  \BibitemOpen
  \bibfield  {author} {\bibinfo {author} {\bibfnamefont {A.}~\bibnamefont {Lucas}},\ }\href@noop {} {\bibfield  {journal} {\bibinfo  {journal} {Frontiers in Physics}\ }\textbf {\bibinfo {volume} {2}},\ \bibinfo {pages} {1} (\bibinfo {year} {2014})}\BibitemShut {NoStop}%
\bibitem [{\citenamefont {Mohseni}, \citenamefont {McMahon},\ and\ \citenamefont {Byrnes}(2022)}]{byrnes2022}%
  \BibitemOpen
  \bibfield  {author} {\bibinfo {author} {\bibfnamefont {N.}~\bibnamefont {Mohseni}}, \bibinfo {author} {\bibfnamefont {P.~L.}\ \bibnamefont {McMahon}}, \ and\ \bibinfo {author} {\bibfnamefont {T.}~\bibnamefont {Byrnes}},\ }\href@noop {} {\bibfield  {journal} {\bibinfo  {journal} {Nature Reviews Physics}\ }\textbf {\bibinfo {volume} {4}},\ \bibinfo {pages} {363} (\bibinfo {year} {2022})}\BibitemShut {NoStop}%
\bibitem [{\citenamefont {Kalinin}\ and\ \citenamefont {Berloff}(2018)}]{kalinin2018}%
  \BibitemOpen
  \bibfield  {author} {\bibinfo {author} {\bibfnamefont {K.}~\bibnamefont {Kalinin}}\ and\ \bibinfo {author} {\bibfnamefont {N.}~\bibnamefont {Berloff}},\ }\href@noop {} {\bibfield  {journal} {\bibinfo  {journal} {Scientific Reports}\ }\textbf {\bibinfo {volume} {8}},\ \bibinfo {pages} {17791} (\bibinfo {year} {2018})}\BibitemShut {NoStop}%
\bibitem [{\citenamefont {Johnson}\ \emph {et~al.}(2011)\citenamefont {Johnson}, \citenamefont {Amin}, \citenamefont {Gildert} \emph {et~al.}}]{Johnson2011}%
  \BibitemOpen
  \bibfield  {author} {\bibinfo {author} {\bibfnamefont {M.}~\bibnamefont {Johnson}}, \bibinfo {author} {\bibfnamefont {M.}~\bibnamefont {Amin}}, \bibinfo {author} {\bibfnamefont {S.}~\bibnamefont {Gildert}},  \emph {et~al.},\ }\href@noop {} {\bibfield  {journal} {\bibinfo  {journal} {Nature}\ ,\ \bibinfo {pages} {194}} (\bibinfo {year} {2011})}\BibitemShut {NoStop}%
\bibitem [{\citenamefont {Ross}\ \emph {et~al.}(2022)\citenamefont {Ross}, \citenamefont {Gradoni}, \citenamefont {Lim},\ and\ \citenamefont {Peng}}]{ross2022}%
  \BibitemOpen
  \bibfield  {author} {\bibinfo {author} {\bibfnamefont {C.}~\bibnamefont {Ross}}, \bibinfo {author} {\bibfnamefont {G.}~\bibnamefont {Gradoni}}, \bibinfo {author} {\bibfnamefont {Q.~J.}\ \bibnamefont {Lim}}, \ and\ \bibinfo {author} {\bibfnamefont {Z.}~\bibnamefont {Peng}},\ }\href {\doibase 10.1109/TAP.2021.3137424} {\bibfield  {journal} {\bibinfo  {journal} {IEEE Transactions on Antennas and Propagation}\ }\textbf {\bibinfo {volume} {70}},\ \bibinfo {pages} {2841} (\bibinfo {year} {2022})}\BibitemShut {NoStop}%
\bibitem [{\citenamefont {Farhi}, \citenamefont {Goldstone},\ and\ \citenamefont {Gutmann}()}]{qaoa}%
  \BibitemOpen
  \bibfield  {author} {\bibinfo {author} {\bibfnamefont {E.}~\bibnamefont {Farhi}}, \bibinfo {author} {\bibfnamefont {J.}~\bibnamefont {Goldstone}}, \ and\ \bibinfo {author} {\bibfnamefont {S.}~\bibnamefont {Gutmann}},\ }\href@noop {} {\bibinfo  {journal} {arXiv:1411.4028}\ }\BibitemShut {NoStop}%
\bibitem [{\citenamefont {Kim}\ \emph {et~al.}(2010)\citenamefont {Kim}, \citenamefont {Chang}, \citenamefont {Korenbilt} \emph {et~al.}}]{kimisingion}%
  \BibitemOpen
\bibfield  {journal} {  }\bibfield  {author} {\bibinfo {author} {\bibfnamefont {K.}~\bibnamefont {Kim}}, \bibinfo {author} {\bibfnamefont {M.}~\bibnamefont {Chang}}, \bibinfo {author} {\bibfnamefont {S.}~\bibnamefont {Korenbilt}},  \emph {et~al.},\ }\href@noop {} {\bibfield  {journal} {\bibinfo  {journal} {Nature}\ }\textbf {\bibinfo {volume} {465}},\ \bibinfo {pages} {590} (\bibinfo {year} {2010})}\BibitemShut {NoStop}%
\bibitem [{\citenamefont {Hamerly}\ \emph {et~al.}(2019)\citenamefont {Hamerly}, \citenamefont {Inagaki}, \citenamefont {McMahon} \emph {et~al.}}]{hamerlyCIM}%
  \BibitemOpen
  \bibfield  {author} {\bibinfo {author} {\bibfnamefont {R.}~\bibnamefont {Hamerly}}, \bibinfo {author} {\bibfnamefont {T.}~\bibnamefont {Inagaki}}, \bibinfo {author} {\bibfnamefont {P.}~\bibnamefont {McMahon}},  \emph {et~al.},\ }\href@noop {} {\bibfield  {journal} {\bibinfo  {journal} {Sci. Adv.}\ }\textbf {\bibinfo {volume} {5}},\ \bibinfo {pages} {1} (\bibinfo {year} {2019})}\BibitemShut {NoStop}%
\bibitem [{\citenamefont {Marandi}\ \emph {et~al.}(2014)\citenamefont {Marandi}, \citenamefont {Wang}, \citenamefont {Takata} \emph {et~al.}}]{marandi2014_parametric_oscillator}%
  \BibitemOpen
  \bibfield  {author} {\bibinfo {author} {\bibfnamefont {A.}~\bibnamefont {Marandi}}, \bibinfo {author} {\bibfnamefont {Z.}~\bibnamefont {Wang}}, \bibinfo {author} {\bibfnamefont {K.}~\bibnamefont {Takata}},  \emph {et~al.},\ }\href@noop {} {\bibfield  {journal} {\bibinfo  {journal} {Nature Photon.}\ }\textbf {\bibinfo {volume} {8}},\ \bibinfo {pages} {937} (\bibinfo {year} {2014})}\BibitemShut {NoStop}%
\bibitem [{\citenamefont {McMahon}\ \emph {et~al.}(2016)\citenamefont {McMahon}, \citenamefont {Marandi}, \citenamefont {Haribara} \emph {et~al.}}]{mcmahon2016_programmable}%
  \BibitemOpen
  \bibfield  {author} {\bibinfo {author} {\bibfnamefont {P.~L.}\ \bibnamefont {McMahon}}, \bibinfo {author} {\bibfnamefont {A.}~\bibnamefont {Marandi}}, \bibinfo {author} {\bibfnamefont {Y.}~\bibnamefont {Haribara}},  \emph {et~al.},\ }\href {\doibase 10.1126/science.aah5178} {\bibfield  {journal} {\bibinfo  {journal} {Science}\ }\textbf {\bibinfo {volume} {354}},\ \bibinfo {pages} {614} (\bibinfo {year} {2016})},\ \Eprint {http://arxiv.org/abs/https://www.science.org/doi/pdf/10.1126/science.aah5178} {https://www.science.org/doi/pdf/10.1126/science.aah5178} \BibitemShut {NoStop}%
\bibitem [{\citenamefont {Inagaki}\ \emph {et~al.}(2016)\citenamefont {Inagaki}, \citenamefont {Haribara}, \citenamefont {Igarashi} \emph {et~al.}}]{inagaki2016}%
  \BibitemOpen
  \bibfield  {author} {\bibinfo {author} {\bibfnamefont {T.}~\bibnamefont {Inagaki}}, \bibinfo {author} {\bibfnamefont {Y.}~\bibnamefont {Haribara}}, \bibinfo {author} {\bibfnamefont {K.}~\bibnamefont {Igarashi}},  \emph {et~al.},\ }\href {\doibase 10.1126/science.aah4243} {\bibfield  {journal} {\bibinfo  {journal} {Science}\ }\textbf {\bibinfo {volume} {354}},\ \bibinfo {pages} {603} (\bibinfo {year} {2016})},\ \Eprint {http://arxiv.org/abs/https://www.science.org/doi/pdf/10.1126/science.aah4243} {https://www.science.org/doi/pdf/10.1126/science.aah4243} \BibitemShut {NoStop}%
\bibitem [{\citenamefont {Calvanese~Strinati}, \citenamefont {Pierangeli},\ and\ \citenamefont {Conti}(2021)}]{strinati2021}%
  \BibitemOpen
  \bibfield  {author} {\bibinfo {author} {\bibfnamefont {M.}~\bibnamefont {Calvanese~Strinati}}, \bibinfo {author} {\bibfnamefont {D.}~\bibnamefont {Pierangeli}}, \ and\ \bibinfo {author} {\bibfnamefont {C.}~\bibnamefont {Conti}},\ }\href {\doibase 10.1103/PhysRevApplied.16.054022} {\bibfield  {journal} {\bibinfo  {journal} {Phys. Rev. Applied}\ }\textbf {\bibinfo {volume} {16}},\ \bibinfo {pages} {054022} (\bibinfo {year} {2021})}\BibitemShut {NoStop}%
\bibitem [{\citenamefont {Ng}\ \emph {et~al.}(2022)\citenamefont {Ng}, \citenamefont {Onodera}, \citenamefont {Kako}, \citenamefont {McMahon}, \citenamefont {Mabuchi},\ and\ \citenamefont {Yamamoto}}]{ng2022}%
  \BibitemOpen
  \bibfield  {author} {\bibinfo {author} {\bibfnamefont {E.}~\bibnamefont {Ng}}, \bibinfo {author} {\bibfnamefont {T.}~\bibnamefont {Onodera}}, \bibinfo {author} {\bibfnamefont {S.}~\bibnamefont {Kako}}, \bibinfo {author} {\bibfnamefont {P.~L.}\ \bibnamefont {McMahon}}, \bibinfo {author} {\bibfnamefont {H.}~\bibnamefont {Mabuchi}}, \ and\ \bibinfo {author} {\bibfnamefont {Y.}~\bibnamefont {Yamamoto}},\ }\href {\doibase 10.1103/PhysRevResearch.4.013009} {\bibfield  {journal} {\bibinfo  {journal} {Phys. Rev. Research}\ }\textbf {\bibinfo {volume} {4}},\ \bibinfo {pages} {013009} (\bibinfo {year} {2022})}\BibitemShut {NoStop}%
\bibitem [{\citenamefont {Vadlamani}, \citenamefont {Xiao},\ and\ \citenamefont {Yablonovitch}(2022{\natexlab{a}})}]{vadlamani2020}%
  \BibitemOpen
  \bibfield  {author} {\bibinfo {author} {\bibfnamefont {S.~K.}\ \bibnamefont {Vadlamani}}, \bibinfo {author} {\bibfnamefont {T.~P.}\ \bibnamefont {Xiao}}, \ and\ \bibinfo {author} {\bibfnamefont {E.}~\bibnamefont {Yablonovitch}},\ }in\ \href {\doibase 10.1109/ICRC57508.2022.00019} {\emph {\bibinfo {booktitle} {2022 IEEE International Conference on Rebooting Computing (ICRC)}}}\ (\bibinfo {year} {2022})\ pp.\ \bibinfo {pages} {45--50}\BibitemShut {NoStop}%
\bibitem [{\citenamefont {Vadlamani}, \citenamefont {Xiao},\ and\ \citenamefont {Yablonovitch}(2022{\natexlab{b}})}]{vadlamani2022}%
  \BibitemOpen
  \bibfield  {author} {\bibinfo {author} {\bibfnamefont {S.~K.}\ \bibnamefont {Vadlamani}}, \bibinfo {author} {\bibfnamefont {T.~P.}\ \bibnamefont {Xiao}}, \ and\ \bibinfo {author} {\bibfnamefont {E.}~\bibnamefont {Yablonovitch}},\ }\href@noop {} {\bibfield  {journal} {\bibinfo  {journal} {https://arxiv.org/abs/2204.02472}\ } (\bibinfo {year} {2022}{\natexlab{b}})}\BibitemShut {NoStop}%
\bibitem [{\citenamefont {Honjo}\ \emph {et~al.}(2021)\citenamefont {Honjo}, \citenamefont {Sonobe}, \citenamefont {Inaba}, \citenamefont {Inagaki}, \citenamefont {Ikuta}, \citenamefont {Yamada}, \citenamefont {Kazama}, \citenamefont {Enbutsu}, \citenamefont {Umeki}, \citenamefont {Kasahara}, \citenamefont {Kawarabayashi},\ and\ \citenamefont {Takesue}}]{cim_honjo}%
  \BibitemOpen
  \bibfield  {author} {\bibinfo {author} {\bibfnamefont {T.}~\bibnamefont {Honjo}}, \bibinfo {author} {\bibfnamefont {T.}~\bibnamefont {Sonobe}}, \bibinfo {author} {\bibfnamefont {K.}~\bibnamefont {Inaba}}, \bibinfo {author} {\bibfnamefont {T.}~\bibnamefont {Inagaki}}, \bibinfo {author} {\bibfnamefont {T.}~\bibnamefont {Ikuta}}, \bibinfo {author} {\bibfnamefont {Y.}~\bibnamefont {Yamada}}, \bibinfo {author} {\bibfnamefont {T.}~\bibnamefont {Kazama}}, \bibinfo {author} {\bibfnamefont {K.}~\bibnamefont {Enbutsu}}, \bibinfo {author} {\bibfnamefont {T.}~\bibnamefont {Umeki}}, \bibinfo {author} {\bibfnamefont {R.}~\bibnamefont {Kasahara}}, \bibinfo {author} {\bibfnamefont {K.}~\bibnamefont {Kawarabayashi}}, \ and\ \bibinfo {author} {\bibfnamefont {H.}~\bibnamefont {Takesue}},\ }\href {\doibase 10.1126/sciadv.abh0952} {\bibfield  {journal} {\bibinfo  {journal} {Science Advances}\ }\textbf {\bibinfo {volume} {7}},\ \bibinfo {pages} {eabh0952} (\bibinfo {year} {2021})}\BibitemShut {NoStop}%
\bibitem [{\citenamefont {Honari-Latifpour}, \citenamefont {Mills},\ and\ \citenamefont {Miri}(2022)}]{latifpour2022}%
  \BibitemOpen
  \bibfield  {author} {\bibinfo {author} {\bibfnamefont {M.}~\bibnamefont {Honari-Latifpour}}, \bibinfo {author} {\bibfnamefont {M.}~\bibnamefont {Mills}}, \ and\ \bibinfo {author} {\bibfnamefont {M.}~\bibnamefont {Miri}},\ }\href@noop {} {\bibfield  {journal} {\bibinfo  {journal} {Commun. Phys.}\ ,\ \bibinfo {pages} {104}} (\bibinfo {year} {2022})}\BibitemShut {NoStop}%
\bibitem [{\citenamefont {Calvanese~Strinati}\ and\ \citenamefont {Conti}(2022)}]{hyperspin}%
  \BibitemOpen
  \bibfield  {author} {\bibinfo {author} {\bibfnamefont {M.}~\bibnamefont {Calvanese~Strinati}}\ and\ \bibinfo {author} {\bibfnamefont {C.}~\bibnamefont {Conti}},\ }\href@noop {} {\bibfield  {journal} {\bibinfo  {journal} {Nat. Commun.}\ ,\ \bibinfo {pages} {7248}} (\bibinfo {year} {2022})}\BibitemShut {NoStop}%
\bibitem [{\citenamefont {Sutton}\ \emph {et~al.}(2017)\citenamefont {Sutton}, \citenamefont {Camsari}, \citenamefont {Behin-Aein},\ and\ \citenamefont {Datta}}]{sutton2017}%
  \BibitemOpen
  \bibfield  {author} {\bibinfo {author} {\bibfnamefont {B.}~\bibnamefont {Sutton}}, \bibinfo {author} {\bibfnamefont {K.~Y.}\ \bibnamefont {Camsari}}, \bibinfo {author} {\bibfnamefont {B.}~\bibnamefont {Behin-Aein}}, \ and\ \bibinfo {author} {\bibfnamefont {S.}~\bibnamefont {Datta}},\ }\href@noop {} {\bibfield  {journal} {\bibinfo  {journal} {Sci. Reports}\ }\textbf {\bibinfo {volume} {7}},\ \bibinfo {pages} {44370} (\bibinfo {year} {2017})}\BibitemShut {NoStop}%
\bibitem [{\citenamefont {Honari-Latifpour}\ and\ \citenamefont {Miri}(2020)}]{Latifpour2020}%
  \BibitemOpen
  \bibfield  {author} {\bibinfo {author} {\bibfnamefont {M.}~\bibnamefont {Honari-Latifpour}}\ and\ \bibinfo {author} {\bibfnamefont {M.-A.}\ \bibnamefont {Miri}},\ }\href {\doibase 10.1103/PhysRevResearch.2.043335} {\bibfield  {journal} {\bibinfo  {journal} {Phys. Rev. Res.}\ }\textbf {\bibinfo {volume} {2}},\ \bibinfo {pages} {043335} (\bibinfo {year} {2020})}\BibitemShut {NoStop}%
\bibitem [{\citenamefont {Parto}\ \emph {et~al.}(2020)\citenamefont {Parto}, \citenamefont {Hayenga}, \citenamefont {Marandi} \emph {et~al.}}]{parto2019}%
  \BibitemOpen
  \bibfield  {author} {\bibinfo {author} {\bibfnamefont {M.}~\bibnamefont {Parto}}, \bibinfo {author} {\bibfnamefont {W.}~\bibnamefont {Hayenga}}, \bibinfo {author} {\bibfnamefont {A.}~\bibnamefont {Marandi}},  \emph {et~al.},\ }\href@noop {} {\bibfield  {journal} {\bibinfo  {journal} {Nat. Mater.}\ }\textbf {\bibinfo {volume} {19}},\ \bibinfo {pages} {725} (\bibinfo {year} {2020})}\BibitemShut {NoStop}%
\bibitem [{\citenamefont {Pierangeli}, \citenamefont {Marcucci},\ and\ \citenamefont {Conti}(2019)}]{pierangeliPRL}%
  \BibitemOpen
  \bibfield  {author} {\bibinfo {author} {\bibfnamefont {D.}~\bibnamefont {Pierangeli}}, \bibinfo {author} {\bibfnamefont {G.}~\bibnamefont {Marcucci}}, \ and\ \bibinfo {author} {\bibfnamefont {C.}~\bibnamefont {Conti}},\ }\href@noop {} {\bibfield  {journal} {\bibinfo  {journal} {Phys. Rev. Lett.}\ }\textbf {\bibinfo {volume} {122}},\ \bibinfo {pages} {213902} (\bibinfo {year} {2019})}\BibitemShut {NoStop}%
\bibitem [{\citenamefont {Pierangeli}, \citenamefont {Marcucci},\ and\ \citenamefont {Conti}(2020)}]{pierangeliadiabatic}%
  \BibitemOpen
  \bibfield  {author} {\bibinfo {author} {\bibfnamefont {D.}~\bibnamefont {Pierangeli}}, \bibinfo {author} {\bibfnamefont {G.}~\bibnamefont {Marcucci}}, \ and\ \bibinfo {author} {\bibfnamefont {C.}~\bibnamefont {Conti}},\ }\href {\doibase 10.1364/OPTICA.398000} {\bibfield  {journal} {\bibinfo  {journal} {Optica}\ }\textbf {\bibinfo {volume} {7}},\ \bibinfo {pages} {1535} (\bibinfo {year} {2020})}\BibitemShut {NoStop}%
\bibitem [{\citenamefont {Fang}, \citenamefont {Huang},\ and\ \citenamefont {Ruan}(2021)}]{Fang2021}%
  \BibitemOpen
  \bibfield  {author} {\bibinfo {author} {\bibfnamefont {Y.}~\bibnamefont {Fang}}, \bibinfo {author} {\bibfnamefont {J.}~\bibnamefont {Huang}}, \ and\ \bibinfo {author} {\bibfnamefont {Z.}~\bibnamefont {Ruan}},\ }\href {\doibase 10.1103/PhysRevLett.127.043902} {\bibfield  {journal} {\bibinfo  {journal} {Phys. Rev. Lett.}\ }\textbf {\bibinfo {volume} {127}},\ \bibinfo {pages} {043902} (\bibinfo {year} {2021})}\BibitemShut {NoStop}%
\bibitem [{\citenamefont {Sun}\ \emph {et~al.}(2022)\citenamefont {Sun}, \citenamefont {Zhang}, \citenamefont {Liu}, \citenamefont {Liu},\ and\ \citenamefont {He}}]{sun2022}%
  \BibitemOpen
  \bibfield  {author} {\bibinfo {author} {\bibfnamefont {W.}~\bibnamefont {Sun}}, \bibinfo {author} {\bibfnamefont {W.}~\bibnamefont {Zhang}}, \bibinfo {author} {\bibfnamefont {Y.}~\bibnamefont {Liu}}, \bibinfo {author} {\bibfnamefont {Q.}~\bibnamefont {Liu}}, \ and\ \bibinfo {author} {\bibfnamefont {Z.}~\bibnamefont {He}},\ }\href {\doibase 10.1364/OL.446789} {\bibfield  {journal} {\bibinfo  {journal} {Opt. Lett.}\ }\textbf {\bibinfo {volume} {47}},\ \bibinfo {pages} {1498} (\bibinfo {year} {2022})}\BibitemShut {NoStop}%
\bibitem [{\citenamefont {Jacucci}\ \emph {et~al.}(2022)\citenamefont {Jacucci}, \citenamefont {Delloye}, \citenamefont {Pierangeli}, \citenamefont {Rafayelyan}, \citenamefont {Conti},\ and\ \citenamefont {Gigan}}]{gigan2022}%
  \BibitemOpen
  \bibfield  {author} {\bibinfo {author} {\bibfnamefont {G.}~\bibnamefont {Jacucci}}, \bibinfo {author} {\bibfnamefont {L.}~\bibnamefont {Delloye}}, \bibinfo {author} {\bibfnamefont {D.}~\bibnamefont {Pierangeli}}, \bibinfo {author} {\bibfnamefont {M.}~\bibnamefont {Rafayelyan}}, \bibinfo {author} {\bibfnamefont {C.}~\bibnamefont {Conti}}, \ and\ \bibinfo {author} {\bibfnamefont {S.}~\bibnamefont {Gigan}},\ }\href@noop {} {\bibfield  {journal} {\bibinfo  {journal} {Phys. Rev. A}\ }\textbf {\bibinfo {volume} {105}},\ \bibinfo {pages} {033502} (\bibinfo {year} {2022})}\BibitemShut {NoStop}%
\bibitem [{\citenamefont {Kumar}, \citenamefont {Zhang},\ and\ \citenamefont {Huang}(2020)}]{kumar2020}%
  \BibitemOpen
  \bibfield  {author} {\bibinfo {author} {\bibfnamefont {S.}~\bibnamefont {Kumar}}, \bibinfo {author} {\bibfnamefont {H.}~\bibnamefont {Zhang}}, \ and\ \bibinfo {author} {\bibfnamefont {Y.}~\bibnamefont {Huang}},\ }\href@noop {} {\bibfield  {journal} {\bibinfo  {journal} {Commun. Phys.}\ ,\ \bibinfo {pages} {108}} (\bibinfo {year} {2020})}\BibitemShut {NoStop}%
\bibitem [{\citenamefont {Ouyang}\ \emph {et~al.}(2024)\citenamefont {Ouyang}, \citenamefont {Liao}, \citenamefont {Feng} \emph {et~al.}}]{ouyang2024}%
  \BibitemOpen
  \bibfield  {author} {\bibinfo {author} {\bibfnamefont {J.}~\bibnamefont {Ouyang}}, \bibinfo {author} {\bibfnamefont {Y.}~\bibnamefont {Liao}}, \bibinfo {author} {\bibfnamefont {X.}~\bibnamefont {Feng}},  \emph {et~al.},\ }\href@noop {} {\bibfield  {journal} {\bibinfo  {journal} {https://arxiv.org/abs/2401.08055}\ } (\bibinfo {year} {2024})}\BibitemShut {NoStop}%
\bibitem [{\citenamefont {Aflatouni}\ \emph {et~al.}(2015)\citenamefont {Aflatouni}, \citenamefont {Abiri}, \citenamefont {Rekhi},\ and\ \citenamefont {Hajimiri}}]{aflounti2015}%
  \BibitemOpen
  \bibfield  {author} {\bibinfo {author} {\bibfnamefont {F.}~\bibnamefont {Aflatouni}}, \bibinfo {author} {\bibfnamefont {B.}~\bibnamefont {Abiri}}, \bibinfo {author} {\bibfnamefont {A.}~\bibnamefont {Rekhi}}, \ and\ \bibinfo {author} {\bibfnamefont {A.}~\bibnamefont {Hajimiri}},\ }\href@noop {} {\bibfield  {journal} {\bibinfo  {journal} {Optics Express}\ }\textbf {\bibinfo {volume} {23}},\ \bibinfo {pages} {21012} (\bibinfo {year} {2015})}\BibitemShut {NoStop}%
\bibitem [{\citenamefont {Chen}\ \emph {et~al.}(2023)\citenamefont {Chen}, \citenamefont {Sludds}, \citenamefont {Davis} \emph {et~al.}}]{vcsel2023}%
  \BibitemOpen
  \bibfield  {author} {\bibinfo {author} {\bibfnamefont {Z.}~\bibnamefont {Chen}}, \bibinfo {author} {\bibfnamefont {A.}~\bibnamefont {Sludds}}, \bibinfo {author} {\bibfnamefont {R.}~\bibnamefont {Davis}},  \emph {et~al.},\ }\href@noop {} {\bibfield  {journal} {\bibinfo  {journal} {Nat. Photon.}\ ,\ \bibinfo {pages} {723}} (\bibinfo {year} {2023})}\BibitemShut {NoStop}%
\bibitem [{\citenamefont {Sun}\ \emph {et~al.}(2013)\citenamefont {Sun}, \citenamefont {Timurdogan}, \citenamefont {Yaacobi}, \citenamefont {Hosseini},\ and\ \citenamefont {Watts}}]{sun2013}%
  \BibitemOpen
  \bibfield  {author} {\bibinfo {author} {\bibfnamefont {J.}~\bibnamefont {Sun}}, \bibinfo {author} {\bibfnamefont {E.}~\bibnamefont {Timurdogan}}, \bibinfo {author} {\bibfnamefont {A.}~\bibnamefont {Yaacobi}}, \bibinfo {author} {\bibfnamefont {E.~S.}\ \bibnamefont {Hosseini}}, \ and\ \bibinfo {author} {\bibfnamefont {M.~R.}\ \bibnamefont {Watts}},\ }\href@noop {} {\bibfield  {journal} {\bibinfo  {journal} {Nature}\ }\textbf {\bibinfo {volume} {493}},\ \bibinfo {pages} {195} (\bibinfo {year} {2013})}\BibitemShut {NoStop}%
\bibitem [{\citenamefont {Poulton}\ \emph {et~al.}(2019)\citenamefont {Poulton}, \citenamefont {Byrd}, \citenamefont {Russo}, \citenamefont {Timurdogan}, \citenamefont {Khandaker}, \citenamefont {Vermeulen},\ and\ \citenamefont {Watts}}]{poulton2019electro}%
  \BibitemOpen
  \bibfield  {author} {\bibinfo {author} {\bibfnamefont {C.~V.}\ \bibnamefont {Poulton}}, \bibinfo {author} {\bibfnamefont {M.~J.}\ \bibnamefont {Byrd}}, \bibinfo {author} {\bibfnamefont {P.}~\bibnamefont {Russo}}, \bibinfo {author} {\bibfnamefont {E.}~\bibnamefont {Timurdogan}}, \bibinfo {author} {\bibfnamefont {M.}~\bibnamefont {Khandaker}}, \bibinfo {author} {\bibfnamefont {D.}~\bibnamefont {Vermeulen}}, \ and\ \bibinfo {author} {\bibfnamefont {M.~R.}\ \bibnamefont {Watts}},\ }\href@noop {} {\bibfield  {journal} {\bibinfo  {journal} {IEEE J. Sel. Top. Quantum Electron.}\ }\textbf {\bibinfo {volume} {29}},\ \bibinfo {pages} {1} (\bibinfo {year} {2019})}\BibitemShut {NoStop}%
\bibitem [{\citenamefont {Xie}\ \emph {et~al.}(2019)\citenamefont {Xie}, \citenamefont {Komljenovic}, \citenamefont {Huang}, \citenamefont {Tran}, \citenamefont {Davenport}, \citenamefont {Torres}, \citenamefont {Pintus},\ and\ \citenamefont {Bowers}}]{xie2019IIIV}%
  \BibitemOpen
  \bibfield  {author} {\bibinfo {author} {\bibfnamefont {W.}~\bibnamefont {Xie}}, \bibinfo {author} {\bibfnamefont {T.}~\bibnamefont {Komljenovic}}, \bibinfo {author} {\bibfnamefont {J.}~\bibnamefont {Huang}}, \bibinfo {author} {\bibfnamefont {M.}~\bibnamefont {Tran}}, \bibinfo {author} {\bibfnamefont {M.}~\bibnamefont {Davenport}}, \bibinfo {author} {\bibfnamefont {A.}~\bibnamefont {Torres}}, \bibinfo {author} {\bibfnamefont {P.}~\bibnamefont {Pintus}}, \ and\ \bibinfo {author} {\bibfnamefont {J.}~\bibnamefont {Bowers}},\ }\href@noop {} {\bibfield  {journal} {\bibinfo  {journal} {Optics Express}\ }\textbf {\bibinfo {volume} {27}},\ \bibinfo {pages} {3642} (\bibinfo {year} {2019})}\BibitemShut {NoStop}%
\bibitem [{\citenamefont {Berloff}\ \emph {et~al.}(2017)\citenamefont {Berloff}, \citenamefont {Silva}, \citenamefont {Kalinin}, \citenamefont {Askitopoulos}, \citenamefont {Topfer}, \citenamefont {Cilibrizzi}, \citenamefont {Langbein},\ and\ \citenamefont {Lagoudakis}}]{berloff2017}%
  \BibitemOpen
  \bibfield  {author} {\bibinfo {author} {\bibfnamefont {N.~G.}\ \bibnamefont {Berloff}}, \bibinfo {author} {\bibfnamefont {M.}~\bibnamefont {Silva}}, \bibinfo {author} {\bibfnamefont {K.}~\bibnamefont {Kalinin}}, \bibinfo {author} {\bibfnamefont {A.}~\bibnamefont {Askitopoulos}}, \bibinfo {author} {\bibfnamefont {J.~D.}\ \bibnamefont {Topfer}}, \bibinfo {author} {\bibfnamefont {P.}~\bibnamefont {Cilibrizzi}}, \bibinfo {author} {\bibfnamefont {W.}~\bibnamefont {Langbein}}, \ and\ \bibinfo {author} {\bibfnamefont {P.~G.}\ \bibnamefont {Lagoudakis}},\ }\href@noop {} {\bibfield  {journal} {\bibinfo  {journal} {Nat. Mater.}\ }\textbf {\bibinfo {volume} {16}},\ \bibinfo {pages} {1120} (\bibinfo {year} {2017})}\BibitemShut {NoStop}%
\bibitem [{\citenamefont {Chalupnik}\ \emph {et~al.}(2023)\citenamefont {Chalupnik}, \citenamefont {Singh}, \citenamefont {Leatham}, \citenamefont {Lon\v{c}ar},\ and\ \citenamefont {Soltani}}]{mvc_phasedarray2023}%
  \BibitemOpen
  \bibfield  {author} {\bibinfo {author} {\bibfnamefont {M.}~\bibnamefont {Chalupnik}}, \bibinfo {author} {\bibfnamefont {A.}~\bibnamefont {Singh}}, \bibinfo {author} {\bibfnamefont {J.}~\bibnamefont {Leatham}}, \bibinfo {author} {\bibfnamefont {M.}~\bibnamefont {Lon\v{c}ar}}, \ and\ \bibinfo {author} {\bibfnamefont {M.}~\bibnamefont {Soltani}},\ }\href@noop {} {\bibfield  {journal} {\bibinfo  {journal} {APL Photonics}\ }\textbf {\bibinfo {volume} {8}},\ \bibinfo {pages} {051305} (\bibinfo {year} {2023})}\BibitemShut {NoStop}%
\bibitem [{\citenamefont {Luo}\ \emph {et~al.}(2023)\citenamefont {Luo}, \citenamefont {Mi}, \citenamefont {Huang},\ and\ \citenamefont {Ruan}}]{luo_multiplexing}%
  \BibitemOpen
  \bibfield  {author} {\bibinfo {author} {\bibfnamefont {L.}~\bibnamefont {Luo}}, \bibinfo {author} {\bibfnamefont {Z.}~\bibnamefont {Mi}}, \bibinfo {author} {\bibfnamefont {J.}~\bibnamefont {Huang}}, \ and\ \bibinfo {author} {\bibfnamefont {Z.}~\bibnamefont {Ruan}},\ }\href {\doibase 10.1126/sciadv.adg6238} {\bibfield  {journal} {\bibinfo  {journal} {Science Advances}\ }\textbf {\bibinfo {volume} {9}},\ \bibinfo {pages} {6238} (\bibinfo {year} {2023})}\BibitemShut {NoStop}%
\bibitem [{\citenamefont {Larocque}\ \emph {et~al.}(2019)\citenamefont {Larocque}, \citenamefont {Ranzani}, \citenamefont {Leatham}, \citenamefont {Tate}, \citenamefont {Niechayev}, \citenamefont {Yengst}, \citenamefont {Komljenovic}, \citenamefont {Fodran}, \citenamefont {Smith},\ and\ \citenamefont {Soltani}}]{hugo2019}%
  \BibitemOpen
  \bibfield  {author} {\bibinfo {author} {\bibfnamefont {H.}~\bibnamefont {Larocque}}, \bibinfo {author} {\bibfnamefont {L.}~\bibnamefont {Ranzani}}, \bibinfo {author} {\bibfnamefont {J.}~\bibnamefont {Leatham}}, \bibinfo {author} {\bibfnamefont {J.}~\bibnamefont {Tate}}, \bibinfo {author} {\bibfnamefont {A.}~\bibnamefont {Niechayev}}, \bibinfo {author} {\bibfnamefont {T.}~\bibnamefont {Yengst}}, \bibinfo {author} {\bibfnamefont {T.}~\bibnamefont {Komljenovic}}, \bibinfo {author} {\bibfnamefont {C.}~\bibnamefont {Fodran}}, \bibinfo {author} {\bibfnamefont {D.}~\bibnamefont {Smith}}, \ and\ \bibinfo {author} {\bibfnamefont {M.}~\bibnamefont {Soltani}},\ }\href@noop {} {\bibfield  {journal} {\bibinfo  {journal} {Optics Express}\ }\textbf {\bibinfo {volume} {27}},\ \bibinfo {pages} {34639} (\bibinfo {year} {2019})}\BibitemShut {NoStop}%
\bibitem [{\citenamefont {Gerchberg}\ and\ \citenamefont {Saxton}(1972)}]{gs}%
  \BibitemOpen
  \bibfield  {author} {\bibinfo {author} {\bibfnamefont {R.~W.}\ \bibnamefont {Gerchberg}}\ and\ \bibinfo {author} {\bibfnamefont {W.~O.}\ \bibnamefont {Saxton}},\ }\href@noop {} {\bibfield  {journal} {\bibinfo  {journal} {Optik}\ }\textbf {\bibinfo {volume} {35}},\ \bibinfo {pages} {237} (\bibinfo {year} {1972})}\BibitemShut {NoStop}%
\bibitem [{\citenamefont {Fienup}(1982)}]{Fienup82}%
  \BibitemOpen
  \bibfield  {author} {\bibinfo {author} {\bibfnamefont {J.~R.}\ \bibnamefont {Fienup}},\ }\href {\doibase 10.1364/AO.21.002758} {\bibfield  {journal} {\bibinfo  {journal} {Appl. Opt.}\ }\textbf {\bibinfo {volume} {21}},\ \bibinfo {pages} {2758} (\bibinfo {year} {1982})}\BibitemShut {NoStop}%
\bibitem [{\citenamefont {Roques-Carmes}\ \emph {et~al.}(2020)\citenamefont {Roques-Carmes}, \citenamefont {Shen}, \citenamefont {Zanoci}, \citenamefont {Prabhu}, \citenamefont {Atieh}, \citenamefont {Jing}, \citenamefont {Dubček}, \citenamefont {Mao}, \citenamefont {Johnson}, \citenamefont {\v{C}eperi\'{c}}, \citenamefont {Joannopoulos}, \citenamefont {Englund},\ and\ \citenamefont {Soljačić}}]{carmes2020}%
  \BibitemOpen
  \bibfield  {author} {\bibinfo {author} {\bibfnamefont {C.}~\bibnamefont {Roques-Carmes}}, \bibinfo {author} {\bibfnamefont {Y.}~\bibnamefont {Shen}}, \bibinfo {author} {\bibfnamefont {C.}~\bibnamefont {Zanoci}}, \bibinfo {author} {\bibfnamefont {M.}~\bibnamefont {Prabhu}}, \bibinfo {author} {\bibfnamefont {F.}~\bibnamefont {Atieh}}, \bibinfo {author} {\bibfnamefont {L.}~\bibnamefont {Jing}}, \bibinfo {author} {\bibfnamefont {T.}~\bibnamefont {Dubček}}, \bibinfo {author} {\bibfnamefont {C.}~\bibnamefont {Mao}}, \bibinfo {author} {\bibfnamefont {M.~R.}\ \bibnamefont {Johnson}}, \bibinfo {author} {\bibfnamefont {V.}~\bibnamefont {\v{C}eperi\'{c}}}, \bibinfo {author} {\bibfnamefont {J.~D.}\ \bibnamefont {Joannopoulos}}, \bibinfo {author} {\bibfnamefont {D.}~\bibnamefont {Englund}}, \ and\ \bibinfo {author} {\bibfnamefont {M.}~\bibnamefont {Soljačić}},\ }\href@noop {} {\bibfield  {journal} {\bibinfo  {journal} {Nat. Comm.}\ }\textbf {\bibinfo {volume} {11}},\ \bibinfo {pages} {249} (\bibinfo {year}
  {2020})}\BibitemShut {NoStop}%
\bibitem [{\citenamefont {Pierangeli}\ \emph {et~al.}(2021)\citenamefont {Pierangeli}, \citenamefont {Marcucci}, \citenamefont {Brunner},\ and\ \citenamefont {Conti}}]{pierangeliPEL}%
  \BibitemOpen
  \bibfield  {author} {\bibinfo {author} {\bibfnamefont {D.}~\bibnamefont {Pierangeli}}, \bibinfo {author} {\bibfnamefont {G.}~\bibnamefont {Marcucci}}, \bibinfo {author} {\bibfnamefont {D.}~\bibnamefont {Brunner}}, \ and\ \bibinfo {author} {\bibfnamefont {C.}~\bibnamefont {Conti}},\ }\href@noop {} {\bibfield  {journal} {\bibinfo  {journal} {Photonics Research}\ }\textbf {\bibinfo {volume} {9}},\ \bibinfo {pages} {1446} (\bibinfo {year} {2021})}\BibitemShut {NoStop}%
\bibitem [{\citenamefont {Skaar}\ and\ \citenamefont {Risvik}(1998)}]{skaar98}%
  \BibitemOpen
  \bibfield  {author} {\bibinfo {author} {\bibfnamefont {J.}~\bibnamefont {Skaar}}\ and\ \bibinfo {author} {\bibfnamefont {K.~M.}\ \bibnamefont {Risvik}},\ }\href {https://opg.optica.org/jlt/abstract.cfm?URI=jlt-16-10-1928} {\bibfield  {journal} {\bibinfo  {journal} {J. Lightwave Technol.}\ }\textbf {\bibinfo {volume} {16}},\ \bibinfo {pages} {1928} (\bibinfo {year} {1998})}\BibitemShut {NoStop}%
\bibitem [{\citenamefont {Katoch}, \citenamefont {Chauhan},\ and\ \citenamefont {Kumar}(2021)}]{katoch2021}%
  \BibitemOpen
  \bibfield  {author} {\bibinfo {author} {\bibfnamefont {S.}~\bibnamefont {Katoch}}, \bibinfo {author} {\bibfnamefont {S.}~\bibnamefont {Chauhan}}, \ and\ \bibinfo {author} {\bibfnamefont {V.}~\bibnamefont {Kumar}},\ }\href@noop {} {\bibfield  {journal} {\bibinfo  {journal} {Multimed Tools Appl}\ ,\ \bibinfo {pages} {8091}} (\bibinfo {year} {2021})}\BibitemShut {NoStop}%
\end{thebibliography}%

\end{document}